\documentclass{article}

\usepackage{microtype}
\usepackage{graphicx}
\usepackage{subfigure}
\usepackage{booktabs} 

\usepackage{xr-hyper}
\usepackage{hyperref}
\usepackage{xr}
\usepackage{multirow}

\usepackage[accepted]{icml2020}

\icmltitlerunning{Full Law Identification in Graphical Models of Missing Data: Completeness Results}

\usepackage{mathtools}
\usepackage{tikz}
\usepackage{amsmath}
\usepackage{amssymb}
\usepackage{multirow}
\usepackage{multicol}
\usepackage{arydshln}
\usepackage{natbib}

\usepackage{amsthm}

\newtheorem{lemma}{Lemma}
\newtheorem{theorem}{Theorem}
\newtheorem{remark}{Remark}

\theoremstyle{definition}

\newenvironment{thma}[1]{\par\noindent{\bf Theorem #1\ }\em}{\em}
\newenvironment{lema}[1]{\par\noindent{\bf Lemma #1\ }\em}{\em}

\newcommand\ci{\perp\!\!\!\perp}
\newcommand{\G}{{\mathcal G}}

\newcommand{\Odds}{\textrm{OR}}

\DeclareMathOperator{\pa}{pa}
\DeclareMathOperator{\ch}{ch}

\DeclareMathOperator{\dis}{dis}
\DeclareMathOperator{\mb}{mb}

\begin{document}

\twocolumn[
\icmltitle{Full Law Identification in Graphical Models of Missing Data: \\ Completeness Results}



\icmlsetsymbol{equal}{*}

\begin{icmlauthorlist}
	\icmlauthor{Razieh Nabi}{equal,jhu}
	\icmlauthor{Rohit Bhattacharya}{equal,jhu}
	\icmlauthor{Ilya Shpitser}{jhu}
\end{icmlauthorlist}

\icmlaffiliation{jhu}{Department of Computer Science, Johns Hopkins University, Baltimore, MD, USA}

\icmlcorrespondingauthor{Razieh Nabi}{rnabi@jhu.edu}
\icmlcorrespondingauthor{Rohit Bhattacharya}{rbhattacharya@jhu.edu}

\icmlkeywords{Missing Data, Identification, Missing Not At Random, Causality, Graphical Models, Selection Bias}

\vskip 0.3in
]



\printAffiliationsAndNotice{\icmlEqualContribution} 

\begin{abstract}
	Missing data has the potential to affect analyses conducted in all fields of scientific study including healthcare, economics, and the social sciences. Several approaches to unbiased inference in the presence of non-ignorable missingness rely on the specification of the target distribution and its missingness process as a probability distribution that factorizes with respect to a directed acyclic graph. In this paper, we address the longstanding question of the characterization of models that are identifiable within this class of missing data distributions. We provide the first completeness result in this field of study --  necessary and sufficient graphical conditions under which, the full  data distribution can be recovered from the observed data distribution. We then simultaneously address issues that may arise due to the presence of both missing data and unmeasured confounding, by extending these graphical conditions and proofs of completeness, to settings where some variables are not just missing, but completely unobserved.
\end{abstract}

\section{Introduction}
\label{sec:intro}

Missing data has the potential to affect analyses conducted in all fields of scientific study, including healthcare, economics, and the social sciences. Strategies to cope with missingness that depends only on the observed data, known as the missing at random (MAR) mechanism, are well-studied \citep{dempster1977maximum,cheng1994nonparametric,robins94estimation,tsiatis06missing}. However, the setting where missingness depends on covariates that may themselves be missing, known as the missing not at random (MNAR) mechanism, is substantially more difficult and under-studied \citep{fielding2008simple, marston2010issues}. MNAR mechanisms are expected to occur quite often in practice, for example, in longitudinal studies with complex patterns of dropout and re-enrollment, or in studies where social stigma may prompt non-response to questions pertaining to drug-use, or sexual activity and orientation, in a way that depends on other imperfectly collected or censored covariates \citep{robins1997non, vansteelandt2007estimation,marra2017simultaneous}.

Previous work on MNAR models has proceeded by imposing a set of restrictions on the full data distribution (the target distribution and its missingness mechanism) that are sufficient to yield identification of the parameter of interest. While there exist MNAR models whose restrictions cannot be represented graphically \citep{tchetgen16discrete}, the restrictions posed in several popular MNAR models such as the permutation model \citep{robins1997non}, the block-sequential MAR model \citep{zhou2010block}, the itemwise conditionally independent nonresponse (ICIN) model \citep{shpitser16consistent, sadinle16itemwise}, and those in \citep{daniel2012using, thoemmes2013selection, martel2013definition, mohan13missing, mohan14missing,  saadati2019adjustment} are either explicitly graphical or can be interpreted as such.

Despite the popularity of graphical modeling approaches for missing data problems, characterization of the class of missing data distributions identified as functionals of the observed data distribution has remained an open question \citep{rozi19mid}. Several algorithms for the identification of the target distribution have been proposed \citep{mohan14missing, shpitser15missing, tian2017recovering, rozi19mid}. We show that even the most general algorithm currently published \citep{rozi19mid} still retains a significant gap in that there exist target distributions that are identified which the algorithm fails to identify. We then present what is, to our knowledge, the first completeness result for missing data models representable as directed acyclic graphs (DAGs) -- a necessary and sufficient graphical condition under which the full data distribution is identified as a function of the observed data distribution. For any given field of study, such a characterization is one of the most powerful results that identification theory can offer, as it comes with the guarantee that if these conditions do not hold, the model is provably not identified.

We further generalize these graphical conditions to settings where some variables are not just missing, but completely unobserved. Such distributions are typically summarized using acyclic directed mixed graphs (ADMGs) \citep{richardson17nested}. We prove, once again, that our graphical criteria are sound and complete for the identification of full laws that are Markov relative to a hidden variable DAG and the resulting summary ADMG. This new result allows us to address two of the most critical issues in practical data analyses simultaneously, those of missingness and unmeasured confounding.

Finally, in the course of proving our results on completeness, we show that the proposed graphical conditions also imply that all missing data models of directed acyclic graphs or acyclic directed mixed graphs that meet these conditions, are in fact sub-models of the MNAR models in \citep{shpitser16consistent, sadinle16itemwise}. This simple, yet powerful result implies that the joint density of these models may be identified using an odds ratio parameterization that also ensures congenial specification of various components of the likelihood \citep{chen07semiparametric, malinsky19noself}. Our results serve as an important precondition for the development of score-based model selection methods for graphical models of missing data, as an alternative to the constraint-based approaches proposed in \citep{strobl2018fast, gain18missing, tu2019causal}, {and directly yield semi-parametric estimators using results in \citep{malinsky19noself}.}

\section{Preliminaries}
\label{sec:prelim}

A directed acyclic graph (DAG) $\G(V)$ consists of a set of nodes $V$ connected through directed edges such that there are no directed cycles. We will abbreviate $\G(V)$ as simply $\G,$ when the vertex set is clear from the given context. Statistical models of a DAG $\G$ are sets of distributions that factorize as $p(V) = \prod_{V_i \in V} p(V_i \mid \pa_\G(V_i))$, where $\pa_\G(V_i)$ are the parents of $V_i$ in $\G$. The absence of edges between variables in $\G,$ relative to a complete DAG entails conditional independence facts in $p(V).$ These can be directly read off from the DAG $\G$ by the well-known d-separation criterion \citep{pearl2009causality}. That is, for disjoint sets $X, Y, Z$, the following \emph{global Markov property} holds: $(X \ci_{\text{d-sep}} Y \mid Z)_\G \implies (X \ci Y \mid Z)_{p(V)}.$ When the context is clear, we will simply use $X \ci Y \mid Z$ to denote the conditional independence between $X$ and $Y$ given $Z.$

In practice, some variables on the DAG may be unmeasured or hidden. In such cases, the distribution $p(V \cup U)$ is Markov relative to a hidden variable DAG $\G(V \cup U),$ where variables in $U$ are unobserved. There may be infinitely many hidden variable DAGs that imply the same set of conditional independences on the observed margin. Hence, it is typical to utilize a single acyclic directed mixed graph (ADMG) consisting of directed and bidirected edges that entails the same set of equality constraints as this infinite class \citep{evans2018margins}. Such an ADMG $\G(V)$ is obtained from a hidden variable DAG $\G(V\cup U)$ via the latent projection operator \cite{verma1990equivalence} as follows. $A \rightarrow B$ exists in $\G(V)$ if there exists a directed path from $A$ to $B$ in $\G(V\cup U)$ with all intermediate vertices in $U.$ An edge $A \leftrightarrow B$ exists in $\G(V)$ if there exists a collider-free path (i.e., there are no consecutive edges of the form $\rightarrow \circ \leftarrow$) from $A$ to $B$ in $\G(V\cup U)$ with all intermediate vertices in $U,$ such that the first edge on the path is an incoming edge into $A$ and the final edge is an incoming edge into $B.$ 

Given a distribution $p(V \cup U)$ that is Markov relative to a hidden variable DAG $\G(V,U),$ conditional independence facts pertaining to the observed margin $p(V)$ can be read off from the ADMG $\G(V)$ by a simple analogue of the d-separation criterion, known as m-separation \citep{richardson2003markov}, that generalizes the notion of a collider to include mixed edges of the form $\rightarrow \circ \leftrightarrow, \leftrightarrow \circ \leftarrow,$ and $\leftrightarrow \circ \leftrightarrow.$

\section{Missing Data Models}
\label{sec:mid}

A missing data model is a set of distributions defined over a set of random variables $\{O, X^{(1)}, R, X\}$, where $O$ denotes the set of variables that are always observed, $X^{(1)}$ denotes the set of variables that are potentially missing, $R$ denotes the set of missingness indicators of the variables in $X^{(1)}$, and $X$ denotes the set of the observed proxies of the variables in $X^{(1)}.$ By definition missingness indicators are binary random variables; however, the state space of variables in $X^{(1)}$ and $O$ are unrestricted. Given $X^{(1)}_i \in X^{(1)}$ and its corresponding missingness indicator $R_i \in R$, the observed proxy $X_i$ is defined as $X_i \equiv X^{(1)}_i$ if $R_i = 1$, and $X_i = ?$ if $R_i = 0$. Hence, $p(X \mid R, X^{(1)})$ is deterministically defined. We call the non-deterministic part of a missing data distribution, i.e, $p(O, X^{(1)}, R)$, the \emph{full law}, and partition it into two pieces: the \emph{target law} $p(O, X^{(1)})$ and the \emph{missingness mechanism} $p(R \mid X^{(1)}, O)$. The censored version of the full law $p(O,R,X),$ that the analyst actually has access to is known as the \emph{observed data distribution}. 

Following the convention in \cite{mohan13missing}, let $\G(V)$ be a missing data DAG, where $V = \{O \cup X^{(1)} \cup R \cup X\}.$ In addition to acyclicity, edges of a missing data DAG are subject to other restrictions: outgoing edges from variables in $R$ cannot point to variables in $\{X^{(1)}, O \}$, each $X_i \in X$ has only two parents in $\G,$ i.e., $R_i$ and $X^{(1)}_i$ (these edges represent the deterministic function above that defines $X_i$, and are shown in gray in all the figures below), and there are no outgoing edges from $X_i$ (i.e., the proxy $X_i$ does not cause any variable on the DAG, however the corresponding full data variable $X^{(1)}_i$ may cause other variables.) A missing data model associated with a missing data DAG $\G$ is the set of distributions $p(O, X^{(1)}, R, X)$ that factorizes as,

{\small
	\begin{align*}
	\prod_{V_i \in O \cup X^{(1)} \cup R} \ p(V_i \mid \pa_\G (V_i)) \prod_{X_i \in X} \ p(X_i \mid X^{(1)}_i, R_i).
	\end{align*}
}%
By standard results on DAG models, conditional independences in $p(X^{(1)},O,R)$ can still be read off from $\G$ by the d-separation criterion \citep{pearl2009causality}. For convenience, we will drop the deterministic terms of the form $p(X_i \mid X^{(1)}_i, R_i)$ from the identification analyses in the following sections since these terms are always identified by construction. 

As an extension, we also consider a hidden variable DAG $\G(V \cup U)$, where $V = \{O, X^{(1)}, R, X\}$ and variables in $U$ are unobserved, to encode missing data models in the presence of unmeasured confounders. In such cases, the full law would obey the nested Markov factorization \citep{richardson17nested} with respect to a missing data ADMG $\G(V)$, obtained by applying the latent projection operator \citep{verma1990equivalence} to the hidden variable DAG $\G(V \cup U).$ As a result of marginalization of latents $U,$ there might exist bi-directed edges (to encode the hidden common causes) between variables in $V$ (bi-directed edges are shown in red in all the figures below). It is straightforward to see that a missing data ADMG obtained via projection of a hidden variable missing data DAG follows the exact same restrictions as stated in the previous paragraph (i.e., no directed cycles, $\pa_\G(X_i) = \{X^{(1)}_i, R_i\}$, every $X_i \in X$ is childless, and there are no outgoing edges from $R_i$ to any variables in $\{X^{(1)}, O\}$.)

\subsection{Identification in Missing Data Models}

The goal of non-parametric identification in missing data models is twofold: identification of the target law $p(O, X^{(1)})$ or functions of it $f(p(O, X^{(1)})),$ and identification of the full law $p(O, X^{(1)}, R),$ in terms of the observed data distribution $p(O, R, X).$  

A compelling reason to study the problem of identification of the full law in and of itself, is due to the fact that many popular methods for model selection or causal discovery, rely on the specification of a well-defined and congenial joint distribution \citep{chickering2002ges, ramsey2015scaling, ogarrio2016hybrid}. A complete theory of the characterization of missing data full laws that are identified opens up the possibility of adapting such methods to settings involving non-ignorable missingness, in order to learn not only substantive relationships between variables of interest in the target distribution, but also the processes that drive their missingness. This is in contrast to previous approaches to model selection under missing data that are restricted to submodels of a single fixed identified model \citep{strobl2018fast, gain18missing, tu2019causal}. Such an assumption may be impractical in complex healthcare settings, for example, where discovering the factors that lead to missingness or study-dropout may be just as important as discovering substantive relations in the underlying data.

Though the focus of this paper is on identification of the full law of missing data models that can be represented by a DAG (or a hidden variable DAG), some of our results naturally extend to identification of the target law (and functionals therein) due to the fact that the target law can be derived from the full law as $\sum_R p(O,X^{(1)},R).$

\begin{remark}
	By chain rule of probability, the target law $p(O, X^{(1)})$ is identified \textit{if and only if} $p(R = 1 \mid O, X^{(1)})$ is identified. The identifying functional is given by 
	{\small
		\begin{align*}
		p(O, X^{(1)}) = \frac{p(O, X^{(1)}, R = 1)}{p(R = 1 \mid O, X^{(1)})}.
		\end{align*}
	}%
	(the numerator is a function of observed data by noting that $X^{(1)} =X$, and is observed when $R = 1$). 
	
	\label{remark_target_law}
\end{remark}

\vspace{0.25cm}
\begin{remark}
	The full law $p(O, X^{(1)}, R)$ is identified \textit{if and only if} $p(R \mid O, X^{(1)})$ is identified. According to Remark~\ref{remark_target_law}, the identifying functional is given by
	{\small
		\begin{align*}
		p(O, X^{(1)}, R) = \frac{p(O, X^{(1)}, R = 1)}{p(R = 1 \mid O, X^{(1)})} \times p(R \mid O, X^{(1)}).
		\end{align*}
	}%
	\label{remark_full_law}
\end{remark}
The rest of the paper is organized as follows. In Section~\ref{sec:incom_mid}, we explain, through examples, why none of the existing identification algorithms put forward in the literature are \emph{complete} in the sense that there exist missing data DAGs whose full law and target law are identified but these algorithms fail to derive an identifying functional for them. In Section \ref{sec:full_law_DAG}, we provide a complete algorithm for full law identification. In Section \ref{sec:full_law_admg}, we further extend our identification results to models where unmeasured confounders are present. We defer all proofs to the Appendix.

\section{Incompleteness of Current Methods}
\label{sec:incom_mid}

In this section, we show that even the most general methods proposed for identification in missing data DAG models remain \emph{incomplete}. In other words, we show that there exist \emph{identified} MNAR models that are representable by DAGs, however all existing algorithms fail to identify both the full and target law for these models. For brevity, we use the procedure proposed in \cite{rozi19mid} as an exemplar. However, as it is the most general procedure in the current literature, failure to identify via this procedure would imply failure by all other existing ones. For each example, we also provide alternate arguments for identification that eventually lead to the general theory  in Sections~\ref{sec:full_law_DAG} and \ref{sec:full_law_admg}.

The algorithm proposed by \cite{rozi19mid} proceeds as follows. For each missingness indicator $R_i,$ the algorithm tries to identify the distribution $p(R_i|\pa_\G(R_i))|_{R=1},$ sometimes referred to as the \emph{propensity score} of $R_i.$ It does so by checking if $R_i$ is conditionally independent (given its parents) of the corresponding missingness indicators of its parents that are potentially missing. If this is the case, the propensity score is identified by a simple conditional independence argument (d-separation). Otherwise, the algorithm checks if this condition holds in post-fixing distributions obtained through recursive application of the \emph{fixing} operator, which roughly corresponds to inverse weighting the current distribution by the propensity score of the variable being fixed \citep{richardson17nested} (a more formal definition is provided in the Appendix.) If the algorithm succeeds in identifying the propensity score for each missingness indicator in this manner, then it succeeds in identifying the target law as Remark~\ref{remark_target_law} suggests, since $p(R=1|O,X^{(1)})=\prod_{R_i\in R}p(R_i|\pa_\G(R))|_{R=1}.$ Additionally, if it is the case that in the course of execution, the propensity score $p(R_i|\pa_\G(R_i))$ for each missingness indicator is also identified at all levels of its parents, then the algorithm also succeeds in identifying the full law (due to Remark~\ref{remark_full_law}).

In order to ground our theory in reality, we now describe a series of hypotheses that may arise during the course of a data analysis that seeks to study the link between the effects of smoking on bronchitis, through the deposition of tar or other particulate matter in the lungs. For each hypothesis, we ask if the investigator is able to evaluate the goodness of fit of the proposed model, typically expressed as a function of the full data likelihood, as a function of just the observed data. In other words, we ask if the full law is identified as a function of the observed data distribution. If it is, this enables the analyst to compare and contrast different hypotheses and select one that fits the data the best.

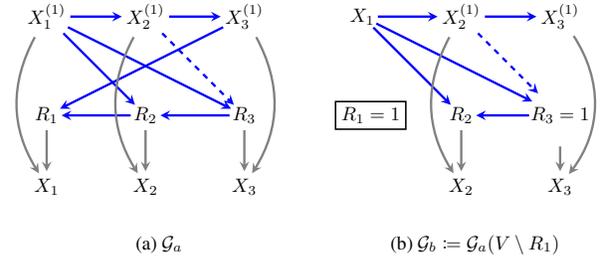
\begin{figure}[t]
	\begin{center}
		\scalebox{0.73}{
			\begin{tikzpicture}[>=stealth, node distance=1.8cm]
			\tikzstyle{format} = [thick, circle, minimum size=1.0mm, inner sep=0pt]
			\tikzstyle{square} = [draw, thick, minimum size=1mm, inner sep=3pt]
			\begin{scope}
			\path[->, very thick]
			node[format] (x11) {$X^{(1)}_1$}
			node[format, right of=x11] (x21) {$X^{(1)}_2$}
			node[format, right of=x21] (x31) {$X^{(1)}_3$}
			node[format, below of=x11] (r1) {$R_1$}
			node[format, below of=x21] (r2) {$R_2$}
			node[format, below of=x31] (r3) {$R_3$}
			node[format, below of=r1, yshift=0.5cm] (x1) {$X_1$}
			node[format, below of=r2, yshift=0.5cm] (x2) {$X_2$}
			node[format, below of=r3, yshift=0.5cm] (x3) {$X_3$}
			(x11) edge[blue] (x21)
			(x21) edge[blue] (x31)
			(r3) edge[blue] (r2)
			(r2) edge[blue] (r1)
			(x11) edge[blue] (r2)
			(x11) edge[blue] (r3)
			(x31) edge[blue] (r1)
			
			(r1) edge[gray] (x1)
			(x11) edge[gray, bend right] (x1)
			(r2) edge[gray] (x2)
			(x21) edge[gray, bend right] (x2)
			(r3) edge[gray] (x3)
			(x31) edge[gray, bend left] (x3)
			(x21) edge[blue, dashed] (r3)
			node [below of=x2, yshift=0.75cm, xshift=0.25cm] {(a) $\G_a$}
			;
			\end{scope}
			\begin{scope}[xshift=5.75cm]
			\path[->, very thick]
			node[format] (x11) {$X_1$}
			node[format, right of=x11] (x21) {$X^{(1)}_2$}
			node[format, right of=x21] (x31) {$X^{(1)}_3$}
			node[square, below of=x11, xshift=0.15cm] (r1) {$R_1=1$}
			node[format, below of=x21] (r2) {$R_2$}
			node[format, below of=x31] (r3) {$R_3 = 1$}
			node[format, below of=r1, yshift=0.5cm] (x1) {}
			node[format, below of=r2, yshift=0.5cm] (x2) {$X_2$}
			node[format, below of=r3, yshift=0.5cm] (x3) {$X_3$}
			(x11) edge[blue] (x21)
			(x21) edge[blue] (x31)
			(r3) edge[blue] (r2)
			(x11) edge[blue] (r2)
			(x11) edge[blue] (r3)
			(x21) edge[blue, dashed] (r3)
			(r2) edge[gray] (x2)
			(x21) edge[gray, bend right] (x2)
			(r3) edge[gray] (x3)
			(x31) edge[gray, bend left=35] (x3)
			node [below of=x2, yshift=0.75cm, xshift=0.25cm] {(b) $\G_b \coloneqq \G_a(V \setminus R_1)$}
			;
			\end{scope}
			\end{tikzpicture}
		}
	\end{center}
	\caption{(a) The missing data DAG used in scenario 1 (without the dashed edge $X^{(1)}_2 \rightarrow R_3$) and scenario 2 (with the dashed edge $X^{(1)}_2 \rightarrow R_3$)  (b) Conditional DAG corresponding to the missing data DAG in (a) after fixing $R_1$, i.e., inverse weighting by the propensity score of $R_1$. }
	\label{fig:fake_sb}
\end{figure}

\textbf{Setup.} To start, the investigator consults a large observational database containing the smoking habits, measurements of particulate matter in the lungs, and results of diagnostic tests for bronchitis on individuals across a city. She notices however, that several entries in the database are missing. This leads her to propose a model like the one shown in Fig.~\ref{fig:fake_sb}(a), where $X_1^{(1)}, X_2^{(1)},$ and $X_3^{(1)}$ correspond to smoking, particulate matter, and bronchitis respectively, and $R_1, R_2,$ and $R_3$ are the corresponding missingness indicators. 

For the target distribution $p(X^{(1)}),$ she proposes a simple mechanism that smoking leads to increased deposits of tar in the lungs, which in turn leads to bronchitis ($X_1^{(1)}\rightarrow X_2^{(1)} \rightarrow X_3^{(1)}$). For the missingness process, she proposes that a suspected diagnosis of bronchitis is likely to lead to an inquiry about the smoking status of the patient ($X_3^{(1)} \rightarrow R_1$), smokers are more likely to get tested for tar and bronchitis ($X_1^{(1)} \rightarrow R_2, X_1^{(1)} \rightarrow R_3$), and ordering a diagnostic test for bronchitis, increases the likelihood of ordering a test for tar, which in turn increases the likelihood of inquiry about smoking status ($R_1 \leftarrow R_2 \leftarrow R_3$).

We now show that for this preliminary hypothesis, if the investigator were to utilize the procedure described in \citep{rozi19mid} she may conclude that it is not possible to identify the full law. We go on to show that such a conclusion would be incorrect, as the full law is, in fact, identified, and provide an alternative means of identification.

\textbf{Scenario 1.}
Consider the missing data DAG model in Fig.~\ref{fig:fake_sb}(a) by excluding the edge $X^{(1)}_2 \rightarrow R_3,$ corresponding to the first hypothesis put forth by the investigator. The propensity score for $R_1$ can be obtained by simple conditioning, noting that $R_1 \ci R_3 \mid X_3^{(1)},R_2$ by d-separation. Hence, $p(R_1 \mid \pa_\G(R_1)) = p(R_1 \mid X_3^{(1)}, R_2) = p(R_1 \mid X_3, R_2, R_3=1).$

Conditioning is not sufficient in order to identify the propensity score for $R_2,$ as $R_2 \not\ci R_1 \mid X_1^{(1)}, R_3$. However, it can be shown that in the distribution $q(V\setminus R_1 \mid R_1=1) \equiv \frac{p(V)}{p(R_1=1 \mid \pa_\G(R_1))}$, $R_2 \ci R_1 \mid X_1, R_3=1$, since this distribution is Markov relative to the graph in Fig.~\ref{fig:fake_sb}(b) (see the Appendix for details). We use the notation $q( \cdot \mid \cdot)$ to indicate that while $q$ acts in most respects as a conditional distribution, it was not obtained from $p(V)$ by a conditioning operation.
This implies that the propensity score for $R_2$ (evaluated at $R = 1$) is identified as $q(R_2 \mid X_1, R_3 = 1, R_1=1).$

Finally, we show that the algorithm in \citep{rozi19mid} is unable to identify the propensity score for $R_3.$ We first note that $R_3 \not\ci R_1 \mid X_1^{(1)}$ in the original problem. Furthermore, as shown in Fig.~\ref{fig:fake_sb}(b), fixing $R_1$ leads to a distribution where $R_3$ is necessarily selected on as the propensity score $p(R_1 \mid \pa_\G(R_1))$ is identified by restricting the data to cases where $R_3=1.$ It is thus impossible to identify the propensity score for $R_3$ in this post-fixing distribution. The same holds if we try to fix $R_2$ as identification of the propensity score for $R_2$ required us to first fix $R_1,$ which we have seen introduces selection bias on $R_3.$

Hence, the procedure in \cite{rozi19mid} fails to identify both the target law and the full law for the problem posed in Fig.~\ref{fig:fake_sb}(a). However, both these distributions are, in fact, identified as we now demonstrate.

A key observation is that even though  the identification of $p(R_3 \mid X^{(1)}_1)$ might not be so straightforward, $p(R_3 \mid X^{(1)}_1, R_2)$ is indeed identified, because by d-separation $R_3 \ci R_1 \mid X^{(1)}_1, R_2$, and therefore $p(R_3 \mid X^{(1)}_1, R_2) = p(R_3 \mid X_1, R_2, R_1 = 1).$ Given that $p(R_3 \mid X^{(1)}_1, R_2)$ and $p(R_2 \mid X^{(1)}_1, R_3=1)$ are both identified (the latter is obtained through as described earlier), we consider exploiting an odds ratio parameterization of the joint density $p(R_2, R_3 \mid \pa_{\G}(R_2, R_3)) = p(R_2, R_3 \mid X^{(1)}_1)$. As we show below, such a parameterization immediately implies the identifiability of this density and consequently, the individual propensity scores for $R_2$ and $R_3$. 

Given disjoint sets of variables $A,B,C$ and reference values $A=a_0, B=b_0,$ the odds ratio parameterization of $p(A, B \mid C)$, given in \cite{chen07semiparametric}, is as follows: 
{\small
	\begin{align}
	& \frac{1}{Z} \times p(A \mid b_0, C) \times p(B \mid a_0, C) \times \text{OR}(A, B \mid C),
	\label{eq:odds_ratio}
	\end{align}	
}%
where 
{\small
	\begin{align*}
	&\text{OR}(A = a, B = b \mid C) \\
	&\hspace{0.5cm} = \frac{p(A = a \mid B = b, C)}{p(A = a_0 \mid B = b, C)} \times \frac{p(A = a_0 \mid B = b_0, C)}{p(A = a \mid B = b_0, C)},
	\end{align*}
}%
and $Z$ is the normalizing term and is equal to 
{\small
	\begin{align*}
	\displaystyle \sum_{A,B} p(A \mid B = b_0, C) \times p(B \mid A = a_0, C) \times \text{OR}(A, B \mid C).
	\end{align*}
}%
Note that ${\small \text{OR}(A, B \mid C) = \text{OR}(B, A \mid C)},$ i.e., the odds ratio is symmetric; see \citep{chen07semiparametric}.

A convenient choice of reference value for the odds ratio in missing data problems is the value $R_i=1.$ Given this reference level and the parameterization of the joint in Eq. ~(\ref{eq:odds_ratio}), we know that
{\small $ p(R_2, R_3 \mid X^{(1)}_1) = 
	\frac{1}{Z} \times p(R_2 \mid R_3 = 1, X^{(1)}_1) \times p(R_3 \mid R_2 = 1, X^{(1)}_1) \times \text{OR}(R_2, R_3 \mid X^{(1)}_1),$} where $Z$ is the normalizing term, and

{\small
	\begin{align*}
	&\text{OR}(R_2 = r_2, R_3 = r_3 \mid X^{(1)}_1) \\
	&\hspace{0.01cm} = \frac{p(R_3 = r_3 \mid R_2 = r_2, X^{(1)}_1)}{p(R_3 = 1 \mid R_2 = r_2, X^{(1)}_1)} \times \frac{p(R_3 = 1 \mid R_2 = 1, X^{(1)}_1)}{p(R_3 = r_3 \mid R_2 = 1, X^{(1)}_1)}. 
	\end{align*}
}%
The conditional pieces $p(R_2 \mid R_3 = 1, X^{(1)}_1)$ and  $p(R_3 \mid R_2 = 1, X^{(1)}_1)$ are already shown to be functions of the observed data. To see that the odds ratio is also a function of observables, recall that $R_3 \ci R_1 \mid R_2, X^{(1)}_1.$ This means that $R_1=1$ can be introduced into each individual piece of the odds ratio functional above, making it so that the entire functional depends only on observed quantities. Since all pieces of the odds ratio parameterization are identified as functions of the observed data, we can conclude that $p(R_2, R_3 \mid X^{(1)}_1)$ is identified as the normalizing term is always identified if all the conditional pieces and the odds ratio are identified. This result, in addition to the fact that $p(R_1 \mid R_2, X^{(1)}_3)$ is identified as before, leads us to the identification of both the target law and the full law, as the missingness process $p(R \mid X^{(1)})$ is identified. 

\textbf{Scenario 2.}
Suppose the investigator is interested in testing an alternate hypothesis to see whether detecting high levels of particulate matter in the lungs, also serves as an indicator to physicians that a diagnostic test for bronchitis should be ordered. This corresponds to the missing data DAG model in Fig.~\ref{fig:fake_sb}(a) by including the edge $X_2^{(1)} \rightarrow R_3.$ Since this is a strict super model of the previous example, the procedure in \cite{rozi19mid} still fails to identify the target and full laws in a similar manner as before. 

However, it is still the case that both the target and full laws are identified. The justification for why the odds ratio parameterization of the joint density $p(R_2,R_3 \mid \pa_{\G}(R_2, R_3)) = p(R_2,R_3 \mid X^{(1)}_1, X^{(1)}_2)$ is identified in this scenario, is more subtle. We have,

{\small
	\begin{align*}
	&p(R_2, R_3 \mid X^{(1)}_1, X^{(1)}_2) = \frac{1}{Z} \times p(R_2 \mid R_3 = 1, X^{(1)}_1, X^{(1)}_2)  \\ &\hspace{0.3cm} \times p(R_3 \mid R_2 = 1, X^{(1)}_1, X^{(1)}_2) \times \text{OR}(R_2, R_3 \mid X^{(1)}_1, X^{(1)}_2).
	\end{align*}
}%

Note that $R_2 \ci X^{(1)}_2 \mid R_3, X^{(1)}_1$, and $R_3 \ci R_1 \mid R_2, X^{(1)}_1, X^{(1)}_2$. Therefore, $p(R_2 \mid R_3 = 1, X^{(1)}_1, X^{(1)}_2) = p(R_2 \mid R_3 = 1, X^{(1)}_1)$ is identified the same way as described in Scenario 1, and $p(R_3 \mid R_2 = 1, X^{(1)}_1, X^{(1)}_2) = p(R_3 \mid R_1 = 1, R_2 = 1, X_1, X_2)$ is a function of the observed data and hence is identified. Now the identification of the joint density $p(R_2,R_3 \mid X^{(1)}_1, X^{(1)}_2)$ boils down to identifiability of the  odds ratio term. By symmetry, we can express the odds ratio $\text{OR}(R_2, R_3 \mid X^{(1)}_1, X^{(1)}_2)$ in two different ways,

{\small
	\begin{align*}
	&\text{OR}(R_2, R_3 \mid X^{(1)}_1, X^{(1)}_2) \\
	&\!\!\!\!= \frac{p(R_2  \mid R_3, X^{(1)}_1)}{p(R_2 = 1 \mid R_3, X^{(1)}_1)} \times \frac{p(R_2 = 1 \mid R_3 = 1, X^{(1)}_1)}{p(R_2 \mid R_3 = 1, X^{(1)}_1)} \\
	&\!\!\!\!= \frac{p(R_3 | R_2, X^{(1)}_1, X^{(1)}_2)}{p(R_3 = 1 |R_2, X^{(1)}_1, X^{(1)}_2)}  \times \frac{p(R_3 = 1 | R_2 = 1, X^{(1)}_1, X^{(1)}_2)}{p(R_3 | R_2 = 1, X^{(1)}_1, X^{(1)}_2)}. 
	\end{align*}
}%

The first equality holds by d-separation ($R_2 \ci X^{(1)}_2 \mid R_3, X^{(1)}_1$). This implies that $\text{OR}(R_2, R_3 \mid X^{(1)}_1, X^{(1)}_2)$ is not a function of $X^{(1)}_2.$ Let us denote this functional by $f_1(R_2, R_3, X^{(1)}_1).$ On the other hand, we can plug-in $R_1 = 1$ to pieces in the second equality since $R_3 \ci R_1 \mid R_2, X^{(1)}_1, X^{(1)}_2$ (by d-separation.) This implies that $\text{OR}(R_2, R_3 \mid X^{(1)}_1, X^{(1)}_2)$ is a function of $X^{(1)}_1$ only through its observed values (i.e. $X_1$). Let us denote this functional by $f_2(R_2, R_3, X_1, X^{(1)}_2, R_1 = 1).$ Since odds ratio is symmetric (by definition), then it must be the case that $f_1(R_2, R_3, X^{(1)}_1) = f_2(R_2, R_3, X_1, X^{(1)}_2, R_1 = 1)$; concluding that $f_2$ cannot be a function of $X^{(1)}_2$, as the left hand side of the equation does not depend on $X^{(1)}_2$. This renders $f_2$ to be a function of only observed quantities, i.e. $f_2 = f_2(R_2, R_3, X_1, R_1 = 1)$. This leads to the conclusion that $p(R_2, R_3 \mid X^{(1)}_1, X^{(1)}_2)$ is identified and consequently the missingness process $p(R \mid X^{(1)})$ in Fig.~\ref{fig:fake_sb}(a) is identified. According to Remarks~\ref{remark_target_law} and \ref{remark_full_law}, both the target and full laws are identified. 

Adding any directed edge to Fig.~\ref{fig:fake_sb}(a) (including the dashed edge) allowed by missing data DAGs results in either a \emph{self-censoring} edge ($X_i^{(1)}\rightarrow R_i$) or a special kind of collider structure called the \emph{colluder} ($X_j^{(1)}\rightarrow R_i \leftarrow R_j$) in \cite{rozi19mid}. We discuss in detail, the link between identification of missing data models of a DAG and the absence of these structures in Section~\ref{sec:full_law_DAG}. 

\textbf{Scenario 3.} 
So far, the investigator has conducted preliminary analyses of the problem while ignoring the issue of unmeasured confounding. In order to address this issue, she first posits an unmeasured confounder $U_1,$ corresponding to genotypic traits that may predispose certain individuals to both smoke and develop bronchitis. She posits another unmeasured confounder $U_2,$ corresponding to the occupation of an individual, that may affect both the deposits of tar found in their lungs (for e.g., construction workers may accumulate more tar than an accountant due to occupational hazards) as well as limit an individual's access to proper healthcare, leading to the absence of a diagnostic test for bronchitis.

\begin{figure}[t]
	\begin{center}
		\scalebox{0.73}{
			\begin{tikzpicture}[>=stealth, node distance=1.8cm]
			\tikzstyle{format} = [thick, circle, minimum size=1.0mm, inner sep=0pt]
			\tikzstyle{square} = [draw, thick, minimum size=1mm, inner sep=3pt]
			\begin{scope}
			\path[->, very thick]
			node[format] (x11) {$X^{(1)}_1$}
			node[format, right of=x11] (x21) {$X^{(1)}_2$}
			node[format, right of=x21] (x31) {$X^{(1)}_3$}
			node[format, below of=x11] (r1) {$R_1$}
			node[format, below of=x21] (r2) {$R_2$}
			node[format, below of=x31] (r3) {$R_3$}
			node[format, below of=r1, yshift=0.5cm] (x1) {$X_1$}
			node[format, below of=r2, yshift=0.5cm] (x2) {$X_2$}
			node[format, below of=r3, yshift=0.5cm] (x3) {$X_3$}
			node[format, above of=x11, yshift=-0.5cm] (u1) {$U_1$}
			node[format, above of=x21, yshift=-0.5cm] (u2) {$U_3$}
			node[format, above of=x31, yshift=-0.5cm] (u3) {$U_2$}
			
			(x11) edge[blue] (x21)
			(x21) edge[blue] (x31)
			(r3) edge[blue] (r2)
			(r2) edge[blue] (r1)
			(x11) edge[blue] (r2)
			(x11) edge[blue] (r3)
			(x31) edge[blue] (r1)
			
			(r1) edge[gray] (x1)
			(x11) edge[gray, bend right] (x1)
			(r2) edge[gray] (x2)
			(x21) edge[blue] (r3)
			(x21) edge[gray, bend left=35] (x2)
			(r3) edge[gray] (x3)
			(x31) edge[gray, bend left] (x3)
			
			(u1) edge[blue] (x31)
			(u1) edge[blue] (x11)
			(u2) edge[blue, dashed] (r1)
			(u2) edge[blue, dashed] (r3)
			(u3) edge[blue] (x21)
			(u3) edge[blue, bend left=40] (r3)
			
			node [below of=x2, yshift=0.75cm, xshift=0.25cm] {(a) $\G(V, U)$}
			;
			\end{scope}
			\begin{scope}[xshift=5.75cm]
			\path[->, very thick]
			node[format] (x11) {$X^{(1)}_1$}
			node[format, right of=x11] (x21) {$X^{(1)}_2$}
			node[format, right of=x21] (x31) {$X^{(1)}_3$}
			node[format, below of=x11] (r1) {$R_1$}
			node[format, below of=x21] (r2) {$R_2$}
			node[format, below of=x31] (r3) {$R_3$}
			node[format, below of=r1, yshift=0.5cm] (x1) {$X_1$}
			node[format, below of=r2, yshift=0.5cm] (x2) {$X_2$}
			node[format, below of=r3, yshift=0.5cm] (x3) {$X_3$}
			(x11) edge[blue] (x21)
			(x11) edge[red, <->, bend left=40] (x31)
			(x21) edge[blue] (x31)
			(r3) edge[blue] (r2)
			(r2) edge[blue] (r1)
			(x11) edge[blue] (r2)
			(x21) edge[blue] (r3)
			(x21) edge[red, <->, bend left] (r3)
			(x11) edge[blue] (r3)
			(x31) edge[blue] (r1)
			(r1) edge[red, <->, bend right, dashed] (r3)
			
			(r1) edge[gray] (x1)
			(x11) edge[gray, bend right] (x1)
			(r2) edge[gray] (x2)
			(x21) edge[gray, bend left=35] (x2)
			(r3) edge[gray] (x3)
			(x31) edge[gray, bend left] (x3)
			node [below of=x2, yshift=0.75cm, xshift=0.25cm] {(b) $\G(V)$}
			;
			\end{scope}
			\end{tikzpicture}
		}
	\end{center}
	\caption{(a) The missing data DAG with unobserved confounders used in scenario 3 (without the dashed edges) and scenario 4 (with the dashed edges). (b) The corresponding missing data ADMGs obtained by applying the latent projection rules to the hidden variable DAG in (a).}
	\label{fig:mnar_admg}
\end{figure}
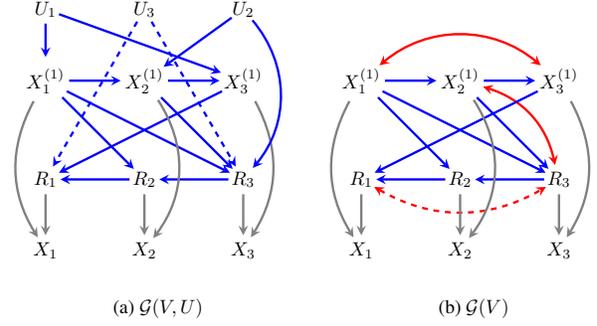

The missing data DAG with unmeasured confounders, corresponding to the aforementioned hypothesis is shown in Fig.~\ref{fig:mnar_admg}(a) (excluding the dashed edges). The corresponding missing data ADMG, obtained by latent projection is shown in Fig.~\ref{fig:mnar_admg}(b) (excluding the dashed bidirected edge). A procedure to identify the full law of such an MNAR model, that is nested Markov with respect to a missing data ADMG, is absent from the current literature. The question that arises, is whether it is possible to adapt the odds ratio parameterization from the previous scenarios, to this setting.

We first note that by application of the chain rule of probability and Markov restrictions, the missingness mechanism still factorizes in the same way as in Scenario 2, i.e., $p(R \mid X^{(1)})=p(R_1 \mid R_2, X^{(1)}_3)\times p(R_2, R_3 \mid X^{(1)}_1, X^{(1)}_2)$ \citep{tian02general}. Despite the addition of the bidirected edges $X_1^{(1)} \leftrightarrow X_3^{(1)}$ and $X_2^{(1)} \leftrightarrow R_3,$ corresponding to unmeasured confounding, it is easy to see that the propensity score for $R_1$ is still identified via simple conditioning. That is, $p(R_1 \mid \pa_\G(R_1)) = p(R_1 \mid X_3, R_2, R_3=1)$ as $R_1 \ci R_3 \mid X_3^{(1)},R_2$ by m-separation. Furthermore, it can also be shown that the two key conditional independences that were exploited in the odds ratio parameterization of $p(R_2,R_3 \mid X^{(1)}),$ still hold in the presence of these additional edges. In particular, $R_2 \ci X_2^{(1)} \mid R_3, X_1^{(1)}$, and $R_3 \ci R_1 \mid R_2, X_1^{(1)}, X_2^{(1)},$ by m-separation. Thus, the same odds ratio parameterization used for identification of the full law in Scenario 2, is also valid for Scenario 3. The full odds ratio parameterization of the MNAR models in Scenarios 2 and 3 is provided in Appendix~B.

\textbf{Scenario 4.} 
Finally, the investigator notices that a disproportionate number of missing entries for smoking status and diagnosis of bronchitis, correspond to individuals from certain neighborhoods in the city. She posits that such missingness may be explained by systematic biases in the healthcare system, where certain ethnic minorities may not be treated with the same level of care. This corresponds to adding a third unmeasured confounder $U_3,$ which affects the ordering of a diagnostic test for bronchitis as well as inquiry about smoking habits, as shown in Fig.~\ref{fig:mnar_admg}(a) (including the dashed edges.) The corresponding missing data ADMG is shown in  Fig.~\ref{fig:mnar_admg}(b) (including the bidirected dashed edge.) Once again, we investigate if the full law is identified, in the presence of an additional unmeasured confounder $U_3$, and the corresponding bidirected edge $R_1 \leftrightarrow R_3.$ 

The missingness mechanism $p(R \mid X^{(1)})$ in Fig.~\ref{fig:mnar_admg}(b) (including the dashed edge) no longer follows the same factorization as the one described in Scenarios 2 and 3, due to the presence of a direct connection between $R_1$ and $R_3.$ According to \cite{tian02general}, this factorization is given as $p(R \mid X^{(1)}) = p(R_1 \mid R_2, R_3, X^{(1)}_1, X^{(1)}_2, X^{(1)}_3) \times p(R_2 \mid R_3, X^{(1)}_1) \times p(R_3 \mid X^{(1)}_1, X^{(1)}_2)$. Unlike the previous scenarios, the propensity score of $R_1$, $p(R_1 \mid R_2, R_3, X^{(1)}_1, X^{(1)}_2, X^{(1)}_3)$, includes $X^{(1)}_1, X^{(1)}_2,$ and $R_3$ past the conditioning bar. Thus, the propensity score of $R_1$ seems to be not identified, since there is no clear way of breaking down the dependency between $R_1$ and $X^{(1)}_1.$ The problematic structure is the path $X^{(1)}_1 \rightarrow R_3 \leftrightarrow R_1$ which contains a collider at $R_3$ that opens up when we condition on $R_3$ in the propensity score of $R_1.$ 

In light of the discussion in previous scenarios, another possibility for identifying $p(R \mid X^{(1)})$ is through analysis of the odds ratio parameterization of the entire missingness mechanism. In Section \ref{sec:full_law_DAG}, we provide a description of the general odds ratio parameterization on an arbitrary number of missingness indicators. For brevity, we avoid re-writing the formula here. We simply point out that the first step in identifying the missingness mechanism via the odds ratio parameterization is arguing whether conditional densities of the form $p(R_i \mid R\setminus R_i = 1, X^{(1)})$ are identified, which is true if $R_i \ci X^{(1)}_i \mid R\setminus R_i, X^{(1)} \setminus X^{(1)}_i.$ 

Such independencies do not hold in Fig.~\ref{fig:mnar_admg}(b) (including the dashed edge) for any of the $R$s, since there exist collider paths between every pair $(X^{(1)}_i, R_i)$ that render the two variables dependent when we condition on everything outside $X^{(1)}_i, R_i$ (by m-separation). Examples of such paths are $X^{(1)}_1 \rightarrow R_3 \leftrightarrow R_1$ and $X^{(1)}_2 \leftrightarrow R_3 \leftrightarrow R_1 \leftarrow R_2$ and $X^{(1)}_3 \rightarrow R_1 \leftrightarrow R_3$. 

In Section~\ref{sec:full_law_admg}, we show that the structures arising in the missing data ADMG presented in Fig.~\ref{fig:mnar_admg}(b) (including the dashed edge), give rise to MNAR models that are provably not identified without further assumptions.

\section{Full Law Identification in DAGs}
\label{sec:full_law_DAG}

\citep{rozi19mid} proved that two graphical structures, namely the self-censoring edge ($X_i^{(1)} \rightarrow R_i$) and the colluder ($X_j^{(1)}\rightarrow R_i \leftarrow R_j$), prevent the identification of full laws in missing data models of a DAG. In this section we exploit an odds ratio parameterization of the missing data process to prove that these two structures are, in fact, the \emph{only} structures that prevent identification, thus yielding a complete characterization of identification for the full law in missing data DAG models. 

We formally introduce the odds ratio parameterization of the missing data process  introduced in \cite{chen07semiparametric}, as a more general version of the simpler form mentioned earlier in Eq.~(\ref{eq:odds_ratio}). Assuming we have $K$ missingness indicators, $p(R \mid X^{(1)}, O)$ can be expressed as follows.

{\small
	\begin{align}
	p(R \mid X^{(1)}, O) =& 
	\ \frac{1}{Z}\times \prod_{k = 1}^{K} \ p(R_k \mid R_{-k} = 1,X^{(1)}, O)  \nonumber \\ 
	\times& \prod_{k = 2}^{K} \text{OR}(R_k, R_{\prec k} \mid R_{\succ k} = 1, X^{(1)}, O), 
	\label{eq:odds_ratio_chen}
	\end{align}
}%
where $R_{-k} = R \setminus R_k, R_{\prec k} = \{R_1, \ldots, R_{k - 1}\}, R_{\succ k} = \{R_{k+1}, \ldots, R_K\}$, and 
{\small
	\begin{align*}
	&\text{OR}(R_k, R_{\prec k} \mid R_{\succ k} = 1, X^{(1)}, O)  \\
	&\hspace{1.cm} = \frac{p(R_k \mid R_{\succ k} = 1, R_{\prec k}, X^{(1)}, O)}{p(R_k = 1 \mid R_{\succ k} = 1, R_{\prec k}, X^{(1)}, O)} \\
	&\hspace{1.5cm} \times 
	\frac{p(R_k = 1 \mid R_{-k} =1, X^{(1)}, O)}{p(R_k \mid R_{-k} = 1, X^{(1)}, O)}.
	\end{align*}
}%
$Z$ in Eq.~(\ref{eq:odds_ratio_chen}) is the normalizing term and is equal to {\small$ \sum_{r} \{ \prod_{k = 1}^{K} \ p(r_k \mid R_{-k} = 1,X^{(1)}, O) \times \prod_{k = 2}^{K} \text{OR}(r_k, r_{\prec k} \mid R_{\succ k} = 1, X^{(1)}, O)\}$}.

Using the odds ratio reparameterization given in Eq.~(\ref{eq:odds_ratio_chen}), we now show that under a standard \textit{positivity assumption}, stating that $p(R  \mid X^{(1)}, O) > \delta > 0$, with probability one for some constant $\delta$, the full law $p(R, X^{(1)}, O)$ of a missing data DAG is identified in the absence of self-censoring edges and colluders.  Moreover, if any of these conditions are violated, the full law is no longer identified. We formalize this result below.

\begin{theorem}
	A full law $p(R, X^{(1)}, O)$ that is Markov relative to a missing data DAG $\G$ is identified if $\G$ does not contain edges of the form $X_i^{(1)} \rightarrow R_i$ (no self-censoring) and structures of the form $X_j^{(1)} \rightarrow R_i \leftarrow R_j$ (no colluders), and the stated positivity assumption holds. Moreover, the resulting identifying functional for the missingness mechanism $p(R \mid X^{(1)}, O)$ is given by the odds ratio parameterization provided in Eq.~\ref{eq:odds_ratio_chen}, and the identifying functionals for the target law and full law are given by Remarks~\ref{remark_target_law} and \ref{remark_full_law}. 
	\label{theorem:full_law_dag}
\end{theorem}

In what follows, we show that the identification theory that we have proposed for the full law in missing data models of a DAG is \emph{sound} and \emph{complete}. Soundness implies that when our procedure succeeds, the model is in fact identified, and the identifying functional is correct. Completeness implies that when our procedure fails, the model is \emph{provably} not identified (non-parametrically). These two properties allow us to derive a precise boundary for what is and is not identified in the space of missing data models that can be represented by a DAG. 

\begin{theorem}
	The graphical condition of no self-censoring and no colluders, put forward in Theorem~\ref{theorem:full_law_dag}, is sound and complete for the identification of full laws $p(R, O, X^{(1)})$ that are Markov relative to a missing data DAG $\G.$
	\label{theorem:sound_complete_dag}
\end{theorem}

We now state an important result that draws a connection between missing data models of a DAG $\G$ that are devoid of self-censoring and colluders, and the itemwise conditionally independent nonresponse (ICIN) model described in \citep{shpitser16consistent, sadinle16itemwise}. As a substantive model, the ICIN model implies that no partially observed variable directly determines its own missingness, and is defined by the restrictions that for every pair $X_i^{(1)}, R_i,$ it is the case that $X_i^{(1)} \ci R_i \mid R_{-i}, X_{-i}^{(1)}, O.$ We utilize this result in the course of proving Theorem~\ref{theorem:sound_complete_dag}.

\begin{lemma}
	A missing data model of a DAG $\G$ that contains no self-censoring edges and no colluders, is a submodel of the ICIN model.
	\label{lem:chain_submodel_dag}
\end{lemma}

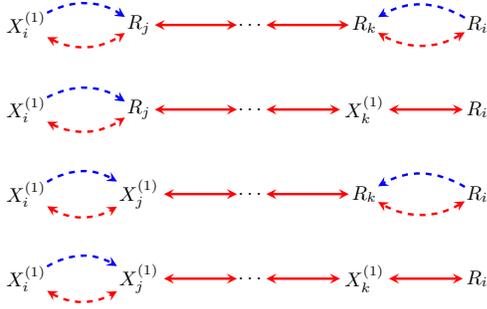
\begin{figure}
	\begin{center}
		\scalebox{0.75}{
			\begin{tikzpicture}[>=stealth, node distance=2cm]
			\tikzstyle{format} = [thick, circle, minimum size=1.0mm, inner sep=0pt]
			\tikzstyle{square} = [draw, thick, minimum size=1mm, inner sep=3pt]	
			\begin{scope}[xshift=0cm]
			\path[->, very thick]
			node[format] (xi) {$X^{(1)}_i$}
			node[format, right of=xi] (rj) {$R_j$}
			node[format, right of=rj] (dot) {$\cdots$}
			node[format, right of=dot] (rk) {$R_k$}
			node[format, right of=rk] (ri) {$R_i$}
			
			(xi) edge[blue, dashed, bend left] (rj)
			(xi) edge[red, dashed, <->, bend right] (rj)
			(rj) edge[red, <->] (dot)
			(dot) edge[red, <->] (rk)
			(ri) edge[blue, dashed, bend right] (rk)
			(ri) edge[red, dashed, <->, bend left] (rk)
			;
			\end{scope}
			\begin{scope}[xshift=0cm, yshift=-1.5cm]
			\path[->, very thick]
			node[format] (xi) {$X^{(1)}_i$}
			node[format, right of=xi] (rj) {$R_j$}
			node[format, right of=rj] (dot) {$\cdots$}
			node[format, right of=dot] (rk) {$X^{(1)}_k$}
			node[format, right of=rk] (ri) {$R_i$}
			
			(xi) edge[blue, dashed, bend left] (rj)
			(xi) edge[red, dashed, <->, bend right] (rj)
			(rj) edge[red, <->] (dot)
			(dot) edge[red, <->] (rk)
			(ri) edge[red, <->] (rk)
			;
			\end{scope}
			\begin{scope}[xshift=0cm, yshift=-3cm]
			\path[->, very thick]
			node[format] (xi) {$X^{(1)}_i$}
			node[format, right of=xi] (xj) {$X^{(1)}_j$}
			node[format, right of=xj] (dot) {$\cdots$}
			node[format, right of=dot] (rk) {$R_k$}
			node[format, right of=rk] (ri) {$R_i$}
			
			(xi) edge[blue, dashed, bend left] (xj)
			(xi) edge[red, dashed, <->, bend right] (xj)
			(xj) edge[red, <->] (dot)
			(dot) edge[red, <->] (rk)
			(ri) edge[blue, dashed, bend right] (rk)
			(ri) edge[red, dashed, <->, bend left] (rk)
			;
			\end{scope}
			\begin{scope}[xshift=0cm, yshift=-4.5cm]
			\path[->, very thick]
			node[format] (xi) {$X^{(1)}_i$}
			node[format, right of=xi] (xj) {$X^{(1)}_j$}
			node[format, right of=xj] (dot) {$\cdots$}
			node[format, right of=dot] (rk) {$X^{(1)}_k$}
			node[format, right of=rk] (ri) {$R_i$}
			
			(xi) edge[blue, dashed, bend left] (xj)
			(xi) edge[red, dashed, <->, bend right] (xj)
			(xj) edge[red, <->] (dot)
			(dot) edge[red, <->] (rk)
			(ri) edge[red, <->] (rk)
			;
			\end{scope}
			\end{tikzpicture}
		}
	\end{center}
	\caption{All possible colluding paths between $X_i^{(1)}$ and $R_i.$ Each pair of dashed edges imply that the presence of either (or both) result in formation of a colluding path.}
	\label{fig:colluder_admg}
\end{figure}

\section{Full Law Identification in the Presence of Unmeasured Confounders}
\label{sec:full_law_admg}

We now generalize identification theory of the full law to scenarios where some variables are not just missing, but completely unobserved, corresponding to the issues faced by the analyst in Scenarios 3 and 4 of Section~\ref{sec:incom_mid}. That is, we shift our focus to the identification of full data laws that are (nested) Markov with respect to a missing data ADMG $\G.$

Previously, we exploited the fact that the absence of colluders and self-censoring edges in a missing data DAG $\G,$ imply a set of conditional independence restrictions of the form $X^{(1)}_i \ci R_i \mid R_{-i}, X^{(1)}_{-i}, O$, for any pair $X^{(1)}_i \in X^{(1)}$ and $R_i \in R$  (see Lemma~\ref{lem:chain_submodel_dag}). We now describe necessary and sufficient graphical conditions that must hold in a missing data ADMG $\G$ to imply this same set of conditional independences. Going forward, we ignore (without loss of generality), the deterministic factors $p(X \mid X^{(1)}, R),$ and the corresponding deterministic edges in $\G,$ in the process of defining this graphical criterion.

A \emph{colliding path} between two vertices $A$ and $B$ is a path on which every non-endpoint node is a collider. We adopt the convention that $A \rightarrow B$ and $A \leftrightarrow B$ are trivially collider paths. We say there exists a \emph{colluding path} between the pair $(X^{(1)}_i, R_i)$ if $X^{(1)}_i$ and $R_i$ are connected through at least one non-deterministic colliding path i.e., one which does not pass through (using deterministic edges) variables in $X.$

We enumerate all possible colluding paths between a vertex $X_i^{(1)}$ and its corresponding missingness indicator $R_i$ in Fig.~\ref{fig:colluder_admg}. Note that both the self-censoring structure and the colluding structure  introduced in \cite{rozi19mid} are special cases of a colluding path. Using the m-separation criterion for ADMGs, it is possible to show that a missing data model of an ADMG $\G$ that contains no colluding paths of the form shown in Fig.~\ref{fig:colluder_admg}, is also a submodel of the ICIN model in \citep{shpitser16consistent, sadinle16itemwise}.

\begin{lemma}
	A missing data model of an ADMG $\G$ that contains no colluding paths is a submodel of the ICIN model.
	\label{lem:chain_submodel_admg}
\end{lemma}

This directly yields a sound criterion for identification of the full law of missing data models of an ADMG $\G$ using the odds ratio parameterization as before.
\begin{theorem}
	A full law $p(R, X^{(1)}, O)$ that is Markov relative to a missing data ADMG $\G$ is identified if  $\G$ does not contain any colluding paths and the stated positivity assumption in Section~\ref{sec:full_law_DAG} holds. Moreover, the resulting identifying functional for the missingness mechanism $p(R \mid X^{(1)}, O)$ is given by the odds ratio parametrization provided in Eq.~\ref{eq:odds_ratio_chen}. 
	\label{theorem:full_law_admg}
\end{theorem}

We now address the question as to whether there exist missing data ADMGs which contain colluding paths but whose full laws are nevertheless identified. We show (see Appendix for proofs), that the presence of a single colluding path of any of the forms shown in Fig.~\ref{fig:colluder_admg}, results in a missing data ADMG $\G$ whose full law $p(X^{(1)}, R, O)$ cannot be identified as a function of the observed data distribution $p(X, R, O).$
\begin{lemma}
	A full law $p(R, X^{(1)}, O)$ that is Markov relative to a missing data ADMG $\G$ containing a colluding path between any pair $X_i^{(1)} \in X^{(1)}$ and $R_i \in R$ is not identified.
	\label{lem:colluding_paths_nonid}
\end{lemma}

Revisiting our example in scenario 4, we note that every $(R_i, X^{(1)}_i)$ pair is connected through at least one colluding path.  Therefore, according to Lemma~\ref{lem:colluding_paths_nonid}, the full law in Fig.~\ref{fig:mnar_admg}(a) including the dashed edge, is not identified. It is worth emphasizing that the existence of at least one colluding path between any pair $(R_i, X^{(1)}_i)$ is sufficient to conclude that the full law is not identified. 

In what follows, we present a result on the soundness and completeness of our graphical condition that represents a powerful unification of non-parametric identification theory in the presence of non-ignorable missingness and unmeasured confounding. To our knowledge, such a result is the first of its kind. We present the theorem below.

\begin{theorem}
	The graphical condition of the absence of colluding paths, put forward in Theorem~\ref{theorem:full_law_admg}, is sound and complete for the identification of full laws $p(X^{(1)}, R, O)$ that are Markov relative to a missing data ADMG $\G.$
	\label{theorem:sound_complete_admg}
\end{theorem}

Throughout the paper, we have focused on identification of the full law which, according to Remark~\ref{remark_target_law}, directly yields identification for the target law. However, identification of the full law is a sufficient but not necessary condition for identification of the target law. In other words, the target law may still be identified despite the presence of colluding paths. Fig.~4(a) in \citep{rozi19mid} is an example of such a case where the full law is not identified due to the colluder structure at $R_2;$ however, as the authors argue the target law remains identified.

\section{Conclusion}
\label{sec:conc}

In this paper, {we concluded an important chapter} in the non-parametric identification theory of missing data models represented via directed acyclic  graphs, possibly in the presence of unmeasured confounders. We provided a simple graphical condition to check if the full law, Markov relative to a (hidden variable) missing data DAG, is identified. We further proved that these criteria are \emph{sound} and \emph{complete}. Moreover, we provided an identifying functional for the missingness process, through an odds ratio parameterization that allows for congenial specification of components of the likelihood. Our results serve as an important precondition for the development of score-based model selection methods that consider a broader class of missing data distributions than the ones considered in prior works. An interesting avenue for future work is exploration of the estimation theory of functionals derived from the identified full data law. To conclude, we note that while identification of the full law is sufficient to identify the target law, there exist identified target laws where the corresponding full law is not identified. We leave a complete characterization of target law identification to future work.

\section*{Acknowledgements}

This project is sponsored in part by the National Science Foundation grant 1939675, the Office of Naval Research grant N00014-18-1-2760, and the Defense Advanced Research Projects Agency under contract HR0011-18-C-0049. The content of the information does not necessarily reflect the position or the policy of the Government, and no official endorsement should be inferred.


\clearpage
\onecolumn
\icmltitle{Supplementary Materials For: \\
	Full Law Identification In Graphical Models Of Missing Data: \\ 
	Completeness Results}

\addtocounter{@affiliationcounter}{-1}

\icmlsetsymbol{equal}{*}

\begin{icmlauthorlist}
	\icmlauthor{Razieh Nabi}{equal,jhu}
	\icmlauthor{Rohit Bhattacharya}{equal,jhu}
	\icmlauthor{Ilya Shpitser}{jhu}
\end{icmlauthorlist}



\icmlkeywords{Missing Data, Identification, Missing Not At Random, Causality, Graphical Models, Selection Bias}

\vskip 0.1in



\printAffiliationsAndNotice{\icmlEqualContribution} 

For clearer presentation of materials in this supplement, we switch to a single-column format. In \textbf{Appendix A}, we provide an overview of the nested Markov model. We summarize the necessary concepts required in order to explain our proof of completeness for identification of the full law in missing data acyclic directed mixed graphs (ADMGs). These concepts draw on the binary parameterization of nested Markov models of an ADMG. In \textbf{Appendix B}, we provide a concrete example of the odds ratio parameterization. In \textbf{Appendix C}, we present proofs that were omitted from the main body of the paper for brevity.

\section*{A. Background: Fixing and Nested Markov Models of an ADMG}
\label{sec:background}

Given a DAG $\G(V \cup U)$ where $U$ contains variables that are unobserved, the \emph{latent projection operator} onto the observed margin produces an acyclic directed mixed graph $\G(V)$ that consists of directed and bidirected edges \citep{verma1990equivalence}. The bidirected connected components of an ADMG $\G(V),$ partition the vertices $V$ into distinct sets known as districts. The district membership of a vertex $V_i$ in $\G$ is denoted $\dis_\G(V_i),$ and the set of all districts in $\G$ is denoted ${\cal D}(\G).$

\cite{evans2018margins} showed that the  nested Markov model \citep{richardson17nested} of an ADMG $\G(V)$ is a smooth super model with fixed dimension, of the underlying latent variable model, that captures all equality constraints and avoids non-regular asymptotics arising from singularities in the parameter space \citep{drton2009discrete, evans2018margins}. We use this fact in order to justify the use of nested Markov models of a missing data ADMG in order to describe full laws that are Markov relative to a missing data DAG with hidden variables. That is, the nested Markov model of a missing data ADMG $\G(V),$ where $V=\{O,X^{(1)}, R, X\},$ is a smooth super model of the missing data DAG model $\G(V\cup U).$ We also utilize nested Markov models of an ADMG $\G(V\setminus X^{(1)}),$ corresponding to projection of the missing data ADMG $\G(V)$ onto variables that are fully observable. While such a model does not capture all equality constraints in the true observed law, it is still a smooth super model of it, thus providing an \emph{upper bound} on the model dimension of the observed law.

\subsection*{CADMGs and Kernels}

The nested Markov factorization of $p(V)$ relative to an ADMG $\G(V)$ is defined with the use of conditional distributions known as \emph{kernels} and their associated \emph{conditional ADMGs} (CADMGs) that are derived from $p(V)$ and $\G(V)$ respectively, via repeated applications of the \emph{fixing operator} \citep{richardson17nested}. A CADMG $\G(V,W),$ is an ADMG whose nodes can be partitioned into random variables $V$ and \emph{fixed} variables $W,$ with the restriction that only outgoing edges may be adjacent to variables in $W.$ A kernel $q_V(V \mid W)$ is a mapping from values in $W$ to normalized densities over $V$ i.e., $\sum_{v\in V}q_V(v\mid w)=1$ \citep{lauritzen96graphical}. Conditioning and marginalization operations in kernels are defined in the usual way.

\subsection*{Fixing and Fixability}

In Section 4 of the main paper, we provided an informal description of fixing as the operation of inverse-weighting by the propensity score of the variable being fixed; we now formalize this notion. A variable $A \in V$ is said to be \emph{fixable} if the paths $A \rightarrow \cdots \rightarrow X$ and $A \leftrightarrow \cdots \leftrightarrow X$ do not both exist for all $X \in V \setminus \{A\}.$ Given a CADMG $\G(V,W)$ where $A$ is fixable, the graphical operator of fixing, denoted $\phi_{A}(\G),$ yields a new CADMG $\G(V\setminus A, W\cup A)$ with all incoming edges into $A$ being removed, and $A$ being set to a fixed value $a.$ Given a kernel $q_V(V\mid W),$ the corresponding probabilistic operation of fixing, denoted $\phi_A(q_V;\G)$ yields a new kernel
\begin{align*}
	q_{V\setminus A}(V \setminus A \mid W \cup A) \equiv \frac{q_{V}(V \mid W)}{q_V(A \mid \mb_\G(A), W)},
\end{align*}
where $\mb_\G(A)$ is the \emph{Markov blanket} of $A,$ defined as the bidirected connected component (district) of $A$ (excluding $A$ itself) and the parents of the district of $A,$ i.e., $\mb_\G(A) \equiv \dis_\G(A) \cup \pa_\G(\dis_\G(A)) \setminus \{A\}.$ It is easy to check that when $\G$ is a DAG, i.e., there are no bidirected edges, the denominator in the probabilistic operation of fixing, reduces to the familiar definition of a simple propensity score.

The notion of fixability can be extended to a set of variables $S\subseteq V$ as follows. A set $S$ is said to be fixable if elements in $S$ can be ordered into a sequence $\sigma_S = \langle S_1, S_2, \dots \rangle$ such that $S_1$ is fixable in $\G,$ $S_2$ is fixable in $\phi_{S_1}(\G),$ and so on. This notion of fixability on sets of variables is essential to the description of the nested Markov model that we present in the following section.

\subsection*{Nested Markov Factorization}

Given a CADMG $\G,$  A set $S \subseteq V$ is said to be \emph{reachable} if there exists a valid sequence of fixing operations on vertices $V\setminus S.$ Further, $S$ is said to be \emph{intrinsic} if it is reachable, and forms a single bidirected connected component or district in $\phi_{\sigma_{V\setminus S}}(\G),$ i.e., the CADMG obtained upon executing all fixing operations given by a valid fixing sequence $\sigma_{V\setminus S}.$

A distribution $p(V)$ is said to obey the nested Markov factorization relative to an ADMG $\G(V)$ if for every fixable set $S,$ and any valid fixing sequence $\sigma_S,$
\begin{align*}
	\phi_{\sigma_S}(p(V);\G) = \prod_{D \in {\cal D}(\phi_{\sigma_S}(\G))} q_D(D \mid \pa_{\phi_{\sigma_S}(\G)}(D)),
\end{align*}
where all kernels appearing in the product above can be constructed by combining kernels corresponding to intrinsic sets i.e., $\{ q_I(I \mid \pa_\G(I)) \mid I \text{ is intrinsic in } \G \}.$
Such a construction is made possible by the fact that all the sets $D$ quantified in the product are districts in a reachable graph derived from $\G.$ 

\cite{richardson17nested} noted that when a distribution $p(V)$ is nested Markov relative to an ADMG $\G,$ all valid fixing sequences yield the same CADMG and kernel so that recursive applications of the fixing operator on a set $V \setminus S$ can simply be denoted as $\phi_{V\setminus S}(\G)$ and $\phi_{V\setminus S}(q_V;\G)$ without explicitly specifying any particular valid order. Thus, the construction of the set of kernels corresponding to intrinsic sets can be characterized as $\{ q_I(I \mid \pa_\G(I)) \mid I \text{ is intrinsic in } \G \} = \{ \phi_{V\setminus I}(p(V; \G)) \mid I \text{ is intrinsic in } \G \},$ and the nested Markov factorization can be re-stated more simply as, for every fixable set $S$ we have,
\begin{align*}
	\phi_S(p(V;\G))=\prod_{D\in {\cal D}\big(\phi_S(\G)\big)} \phi_{V\setminus D}(p(V); \G),
\end{align*}
An important result from \citep{richardson17nested} states that if $p(V\cup U)$ is Markov relative to a DAG $\G(V\cup U),$ then $p(V)$ is nested Markov relative to the ADMG $\G(V)$ obtained by latent projection.

\subsection*{Binary Parameterization of Nested Markov Models}

From the above factorization, it is clear that intrinsic sets given their parents form the atomic units of the nested Markov model. Using this observation, a smooth parameterization of discrete nested Markov models was provided by \cite{evans2014markovian}. We now provide a short description of how to derive the so-called Moebius parameters of a \emph{binary} nested Markov model.

For each district $D \in {\cal D}(\G),$ consider all possible subsets $S \subseteq D.$ If $S$ is intrinsic (that is, reachable and bidirected connected in $\phi_{V\setminus S}(\G)$), define the head $H$ of the intrinsic set to be all vertices in $S$ that are childless in $\phi_{V\setminus S}(\G),$ and the tail $T$ to be all parents of the head in the CADMG $\phi_{V\setminus S}(\G),$ excluding the head itself. More formally, $H \equiv \{V_i \in S \mid \ch_{\phi_{_{V\setminus S}(\G)}}(V_i)=\emptyset\},$ and $T \equiv \pa_{\phi_{_{V\setminus S}(\G)}}(H) \setminus H.$ The corresponding set of Moebius parameters for this intrinsic head and tail pair parameterizes the kernel $q_S(H=0 \mid T),$ i.e., the kernel where all variables outside the intrinsic set $S$ are fixed, and all elements of the head are set to zero given the tail. Note that these parameters are, in general, \emph{variationally dependent} (in contrast to variationally independent in the case of an ordinary DAG model) as the heads and tails in these parameter sets may overlap. The joint density for any query $p(V=v),$ can be obtained through the Moebius inversion formula; see \citep{lauritzen96graphical, evans2014markovian} for details. For brevity, we will denote $q_S(H=0 \mid T)$ as simply $q(H=0 \mid T),$ as it will be clear from the given context what variables are still random in the kernel corresponding to a given intrinsic set.

\subsection*{Binary Parameterization of Missing Data ADMGs}

We use the parameterization described in the previous section in order to count the number of parameters required to parameterize the full law of a missing data ADMG and its corresponding observed law. We then use this to reason that if the number of parameters in the full law exceeds those in the observed law, it is impossible to establish a map from the observed law to the full law. This in turn implies that such a full law is not identified.

The binary parameterization of the \textbf{full law} of a missing data ADMG $\G(X^{(1)},O,R,X)$ is exactly the same as that of an ordinary ADMG, except that the deterministic factors $p(X_i \mid R_i, X_i^{(1)}),$ can be ignored, as $X_i = X_i^{(1)}$ with probability one when $R_i = 1,$ and $X_i = ?$ with probability one when $R_i=0.$ 

The \textbf{observed law} is parameterized as follows. First, variables in $X^{(1)}$ are treated as completely unobserved, and an observed law ADMG $\G(X,O,R)$ is obtained by applying the latent projection operator to $\G(X^{(1)},O,R,X).$ The Moebius parameters are then derived in a similar manner as before, with the additional constraint that if $X_i \in X$ appears in the head of a Moebius parameter, and the corresponding missingness indicator $R_i$ appears in the tail, then the kernel must be restricted to cases where $R_i=1.$ This is because when $R_i=0,$ the probability of the head taking on any value, aside from those where $X_i=?,$ is deterministically defined to be $0.$

Note that parameterizing the observed law by treating variables in $X^{(1)}$ as fully unobserved does not quite capture all equality constraints that may be detectable in the observed law, as these variables are, in fact, sometimes observable when their corresponding missingness indicators are set to one. Indeed, a smooth parameterization of the observed law of missing data models that captures all constraints implied by the model, is still an open problem. Nevertheless, parameterizing an observed law ADMG, such as the one mentioned earlier, provides an \emph{upper bound} on the number of parameters required to parameterize the true observed law. This suffices for our purposes, as demonstrating that the upper bound on the number of parameters in the observed law is less than the number of parameters in the full law, is sufficient to prove that the full law is not identified.

\clearpage
\section*{B. Example: Odds Ratio Parameterization}
\label{sec:odds_ratio_example}

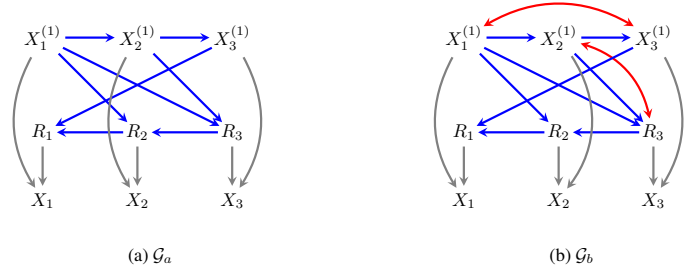
\begin{figure}[t]
	\begin{center}
		\scalebox{0.7}{
			\begin{tikzpicture}[>=stealth, node distance=1.8cm]
			\tikzstyle{format} = [thick, circle, minimum size=1.0mm, inner sep=0pt]
			\tikzstyle{square} = [draw, thick, minimum size=1mm, inner sep=3pt]
			\begin{scope}
			\path[->, very thick]
			node[format] (x11) {$X^{(1)}_1$}
			node[format, right of=x11] (x21) {$X^{(1)}_2$}
			node[format, right of=x21] (x31) {$X^{(1)}_3$}
			node[format, below of=x11] (r1) {$R_1$}
			node[format, below of=x21] (r2) {$R_2$}
			node[format, below of=x31] (r3) {$R_3$}
			node[format, below of=r1, yshift=0.5cm] (x1) {$X_1$}
			node[format, below of=r2, yshift=0.5cm] (x2) {$X_2$}
			node[format, below of=r3, yshift=0.5cm] (x3) {$X_3$}
			(x11) edge[blue] (x21)
			(x21) edge[blue] (x31)
			(r3) edge[blue] (r2)
			(r2) edge[blue] (r1)
			(x11) edge[blue] (r2)
			(x11) edge[blue] (r3)
			(x31) edge[blue] (r1)
			
			(r1) edge[gray] (x1)
			(x11) edge[gray, bend right] (x1)
			(r2) edge[gray] (x2)
			(x21) edge[gray, bend right] (x2)
			(r3) edge[gray] (x3)
			(x31) edge[gray, bend left] (x3)
			(x21) edge[blue] (r3)
			node [below of=x2, yshift=0.75cm, xshift=0.25cm] {(a) $\G_a$}
			;
			\end{scope}
			\begin{scope}[xshift=8cm]
			\path[->, very thick]
			node[format] (x11) {$X^{(1)}_1$}
			node[format, right of=x11] (x21) {$X^{(1)}_2$}
			node[format, right of=x21] (x31) {$X^{(1)}_3$}
			node[format, below of=x11] (r1) {$R_1$}
			node[format, below of=x21] (r2) {$R_2$}
			node[format, below of=x31] (r3) {$R_3$}
			node[format, below of=r1, yshift=0.5cm] (x1) {$X_1$}
			node[format, below of=r2, yshift=0.5cm] (x2) {$X_2$}
			node[format, below of=r3, yshift=0.5cm] (x3) {$X_3$}
			(x11) edge[blue] (x21)
			(x11) edge[red, <->, bend left=30] (x31)
			(x21) edge[blue] (x31)
			(r3) edge[blue] (r2)
			(r2) edge[blue] (r1)
			(x11) edge[blue] (r2)
			(x21) edge[blue] (r3)
			(x21) edge[red, <->, bend left] (r3)
			(x11) edge[blue] (r3)
			(x31) edge[blue] (r1)
			(r1) edge[gray] (x1)
			(x11) edge[gray, bend right] (x1)
			(r2) edge[gray] (x2)
			(x21) edge[gray, bend left=35] (x2)
			(r3) edge[gray] (x3)
			(x31) edge[gray, bend left] (x3)
			node [below of=x2, yshift=0.75cm, xshift=0.25cm] {(b) $\G_b$}
			;
			\end{scope}
			\end{tikzpicture}
		}
	\end{center}
	\caption{(a) The missing data DAG model used in Scenario 2. (b) the missing data ADMG model used in Scenario 3.}
	\label{fig:examples}
\end{figure}

To build up a more concrete intuition for Theorems~\ref{theorem:full_law_dag} and \ref{theorem:full_law_admg}, we provide an example of the odds ratio parameterization for the missing data models used in Scenarios 2 and 3 of the main paper, reproduced here in Figs.~\ref{fig:examples}(a, b). Utilizing the order $R_1, R_2, R_3$ on the missingness indicators, the odds ratio parameterization of the missing data process for both models is as follows.
\begin{align}
	&\frac{1}{Z} \times \bigg(\prod_{k=1}^3 p(R_i \mid R_{-i}=1, X^{(1)})\bigg) \times \text{OR}(R_1, R_2, \mid R_3=1, X^{(1)}) \times \text{OR}(R_3, (R_1, R_2) \mid X^{(1)}).
	\label{eq:example}
\end{align}
We now argue that each piece in Eq.~\ref{eq:example} is identified. Note that, in the missing data DAG shown in Fig.~\ref{fig:examples}(a), $R_i \ci X_i^{(1)} \mid R_{-i}, X_{-i}^{(1)}$ by d-separation. The same is true for the missing data ADMG in Fig.~\ref{fig:examples}(b) by m-separation. Thus, in both cases, the product over conditional pieces of each $R_i$ given the remaining variables is not a function $X_i^{(1)},$ and is thus a function of observed data. We now show that $\text{OR}(R_1, R_2 \mid R_3=1, X^{(1)})$ is not a function of $X_1^{(1)}, X_2^{(1)}$ by utilizing the symmetry property of the odds ratio.
{\small
	\begin{align*}
		\text{OR}(R_1,R_2 \mid R_3=1, X^{(1)}) &= \frac{p(R_1 \mid R_2, R_3=1, X_2^{(1)}, X_3^{(1)})}{p(R_1=1 \mid R_2, R_3=1, X_2^{(1)}, X_3^{(1)})} \times \frac{p(R_1=1 \mid R_2=1, R_3=1, X_2^{(1)}, X_3^{(1)})}{p(R_1 \mid R_2=1, R_3=1, X_2^{(1)}, X_3^{(1)})} \\
		= \text{OR}(R_2,R_1 \mid R_3=1, X^{(1)}) &= \frac{p(R_2 \mid R_1, R_3=1, X_1^{(1)}, X_3^{(1)})}{p(R_2=1 \mid R_1, R_3=1, X_1^{(1)}, X_3^{(1)})} \times \frac{p(R_2=1 \mid R_1=1, R_3=1, X_1^{(1)}, X_3^{(1)})}{p(R_2 \mid R_1=1, R_3=1, X_1^{(1)}, X_3^{(1)})}.
	\end{align*}
}
Thus, from the first equality, the odds ratio is not a function of $X_2^{(1)}$ as $R_1 \ci X_1^{(1)} \mid R_{-1}, X_{-1}^{(1)}$ by d-separation in Fig.~\ref{fig:examples}(a) and by m-separation in Fig.~\ref{fig:examples}(b). A symmetric argument holds for $X_2^{(1)}$ and $R_2$ as seen in the second and third equalities. Hence, the odds ratio is only a function of $X_3^{(1)},$ which is observable, as the function is evaluated at $R_3=1.$

We now utilize an identity from \citep{chen2015likelihood} in order to simplify the final term in Eq.~\ref{eq:example}. That is, 
\begin{align*}
	\text{OR}(R_3, (R_1, R_2) \mid X^{(1)}) &= \text{OR}(R_3, R_2 \mid R_1=1, X^{(1)})\  \text{OR}(R_3, R_1 \mid R_2, X^{(1)}) \\
	&= \text{OR}(R_3, R_2 \mid R_1=1, X^{(1)})\ \text{OR}(R_3, R_1 \mid R_2=1, X^{(1)})\  \underset{f(R_1, R_2, R_3 \mid X^{(1)})}{\underbrace{\frac{\text{OR}(R_3, R_1 \mid R_2, X^{(1)})}{\text{OR}(R_3, R_1 \mid R_2=1, X^{(1)})}}}.
\end{align*}
The first two pairwise odds ratio terms are functions of observed data using an analogous argument that draws on the symmetry property of the odds ratio and the conditional independence $R_i \ci X_i \mid R_{-i}, X_{-i}^{(1)},$ as before. The final term $f(R_1, R_2, R_3 \mid X^{(1)})$, is a three-way interaction term on the odds ratio scale and can be expressed in three different ways as follows \citep{chen2015likelihood},
\begin{align*}
	\frac{\text{OR}(R_3, R_1 \mid R_2, X^{(1)})}{\text{OR}(R_3, R_1 \mid R_2=1, X^{(1)})} = \frac{\text{OR}(R_2, R_3 \mid R_1, X^{(1)})}{\text{OR}(R_2, R_3 \mid R_1=1, X^{(1)})} = \frac{\text{OR}(R_1, R_2 \mid R_3, X^{(1)})}{\text{OR}(R_1, R_2 \mid R_3=1, X^{(1)})}.
\end{align*}
From the first equality, we note by symmetry of the odds ratio and conditional independence that $f$ is not a function of $X^{(1)}_1, X^{(1)}_3$. Similarly, from the second equality, we note that $f$ is not a function of $X^{(1)}_2, X^{(1)}_3$. Finally, from the third equality, we note that $f$ is not a function of $X^{(1)}_1, X^{(1)}_2$. Therefore, $f$ is not a function of $X^{(1)}_1, X^{(1)}_2, X^{(1)}_3$ and is identified. 

The normalizing function $Z,$ is a function of all the pieces that we have already shown to be identified, and is therefore also identified. Thus, the missing data mechanisms $p(R \mid X^{(1)}),$ and consequently, the full laws corresponding to the missing data graphs shown in Figs.~\ref{fig:examples}(a,b) are identified by Remark~\ref{remark_full_law}.

\section*{C. Proofs}
\label{sec:proofs}

We first prove Lemmas~\ref{lem:chain_submodel_dag} and \ref{lem:chain_submodel_admg} as we use them in the course of proving Theorems~\ref{theorem:full_law_dag} and \ref{theorem:full_law_admg}. We start with Lemma~\ref{lem:chain_submodel_admg}, as the proof for Lemma~\ref{lem:chain_submodel_dag} simplifies to a special case.

\begin{lema}{\ref{lem:chain_submodel_admg}}
	A missing data model of an ADMG $\G$ that contains no colluding paths is a submodel of the itemwise conditionally independent nonresponse model described in \citep{shpitser16consistent, sadinle16itemwise}.
\end{lema}
\begin{proof}
	The \emph{complete Markov blanket} of a vertex $V_i$ in an ADMG $\G,$ denoted $\mb_\G^c(V_i)$ is the set of vertices such that $V_i \ci V_{-i} \setminus \mb_\G^c(V_i) \mid \mb_\G^c(V_i)$ \citep{pearl1988probabilistic, richardson2003markov}. In ADMGs, this set corresponds to the Markov blanket of $V_i,$ its children, and the Markov blanket of its children. That is,
	\begin{align*}
		\mb_\G^c(V_i) \equiv \mb_\G(V_i) \cup \bigg( \bigcup_{V_j \in \ch_\G(V_i)} V_j \cup \mb_\G(V_j) \bigg) \setminus \{V_i\}.
	\end{align*}
	
	Without loss of generality, we ignore the part of the graph involving the deterministic factors $p(X \mid X^{(1)}, R)$ and the corresponding deterministic edges, in the construction of the Markov blanket and complete Markov blanket of variables in a missing data graph $\G(X^{(1)},O,R).$ We now show that the absence of non-deterministic colluder paths between a pair $X_i^{(1)}$ and $R_i$ in $\G$ implies that $X_i^{(1)} \notin \mb_\G^c(R_i).$
	
	\begin{itemize}
		\item $X_i^{(1)}$ is not a parent of $R_i,$ as $X_i^{(1)} \rightarrow R_i$ is trivially a colluder path.
		\item $X_i^{(1)}$ is not in the district of $R_i,$ as $X_i^{(1)} \leftrightarrow \cdots \leftrightarrow R_i$ is also a colluder path.
	\end{itemize}
	
	These two points together imply that $X_i^{(1)} \notin \mb_\G(R_i).$ We now show that the union over children of $R_i$ and their Markov blankets also exclude $X_i^{(1)}.$
	\begin{itemize}
		\item $X_i^{(1)}$ is not a child of $R_i,$ as directed edges from $R_i$ to variables in $X^{(1)}$ are ruled out by construction in missing data graphs.
		\item $X_i^{(1)}$ is also not in the district of any children of $R_i,$ as $R_i \rightarrow \cdots \leftrightarrow X_i^{(1)}$ is a colluding path.
		\item $X_i^{(1)}$ is also not a parent of the district of any children of $R_i,$ as $R_i \rightarrow \cdots \leftarrow X_i^{(1)}$ is a colluding path.
	\end{itemize}
	
	These three points together rule out the possibility that $X_i^{(1)}$ is present in the union over children and Markov blankets of children of $R_i.$ Thus, we have shown that $X_i^{(1)} \not\in \mb_\G^c(R_i).$ This implies the following,
	\begin{align*}
		R_i \ci V \setminus \{R_i, \mb_\G^c(R_i) \} \mid \mb_\G^c(R_i) \implies
		R_i \ci X_i^{(1)} \mid \mb_\G^c(R_i).
	\end{align*}
	
	By semi-graphoid axioms (see for example, \citep{lauritzen96graphical, pearl2009causality}) this yields the conditional independence $R_i \ci X_i^{(1)} \mid R_{-i}, X^{(1)}_{-i}, O.$
	
	The same line of reasoning detailed above can be used for all $R_i \in R,$ which then gives us the set of conditional independences implied by the no self-censoring model. That is,
	\begin{align*}
		R_i \ci X^{(1)}_i \mid R_{-i}, X^{(1)}_{-i}, O, \quad  \forall R_i \in R.
	\end{align*}
\end{proof}

\begin{lema}{\ref{lem:chain_submodel_dag}} 
	A missing data model of a DAG $\G$ that contains no self-censoring edges and no colluders, is a submodel of the  itemwise conditionally independent nonresponse model described in \citep{shpitser16consistent, sadinle16itemwise}.
\end{lema}
\begin{proof}
	A DAG is simply a special case of an ADMG with no bidirected edges. Consequently the only two types of colluding paths, are self-censoring edges ($X_i^{(1)} \rightarrow R_i$) and colluder structures ($X_i^{(1)} \rightarrow R_j \leftarrow R_i).$ Thus, the absence of these two structures in a missing data DAG $\G,$ rules out all possible colluding paths. The rest of the proof then carries over straightforwardly from Lemma~\ref{lem:chain_submodel_admg}.
\end{proof}

\begin{thma}{\ref{theorem:full_law_dag}}
	A full law $p(R, X^{(1)}, O)$ that is Markov relative to a missing data DAG $\G$ is identified if $\G$ does not contain edges of the form $X_i^{(1)} \rightarrow R_i$ (no self-censoring) and structures of the form $X_j^{(1)} \rightarrow R_i \leftarrow R_j$ (no colluders), and the stated positivity assumption holds. Moreover, the resulting identifying functional for the missingness mechanism $p(R \mid X^{(1)}, O)$ is given by the odds ratio parameterization provided in Eq.~\ref{eq:odds_ratio_chen} of the main draft, and the identifying functionals for the target law and full law are given by Remarks~\ref{remark_target_law} and \ref{remark_full_law}. 
\end{thma}
\begin{proof}
	Given Eq.~(\ref{eq:odds_ratio_chen}), we know that 
	\begin{align*}
		p(R \mid X^{(1)}, O) =& 
		\ \frac{1}{Z}\times \prod_{k = 1}^{K} \ p(R_k \mid R_{-k} = 1,X^{(1)}, O)  
		\times
		\prod_{k = 2}^{K} \text{OR}(R_k, R_{\prec k} \mid R_{\succ k} = 1, X^{(1)}, O), 
	\end{align*}
	where $R_{-k} = R \setminus R_k, R_{\prec k} = \{R_1, \ldots, R_{k - 1}\}, R_{\succ k} = \{R_{k+1}, \ldots, R_K\}$, and 
	\begin{align*}
		&\text{OR}(R_k, R_{\prec k} \mid R_{\succ k} = 1, X^{(1)}, O)  
		= \frac{p(R_k \mid R_{\succ k} = 1, R_{\prec k}, X^{(1)}, O)}{p(R_k = 1 \mid R_{\succ k} = 1, R_{\prec k}, X^{(1)}, O)} 
		\times 
		\frac{p(R_k = 1 \mid R_{-k} =1, X^{(1)}, O)}{p(R_k \mid R_{-k} = 1, X^{(1)}, O)},
	\end{align*}
	and $Z$ is the normalizing term and is equal to {\small$ \sum_{r} \{ \prod_{k = 1}^{K} \ p(r_k \mid R_{-k} = 1,X^{(1)}, O) \times \prod_{k = 2}^{K} \text{OR}(r_k, r_{\prec k} \mid R_{\succ k} = 1, X^{(1)}, O)\}$}. If we can prove that all the pieces in this factorization are identified, then the missingness process is identified and so is the full law. We provide the proof in two steps. Our proof is similar to the identification proof of the no self-censoring model given in  \cite{malinsky19noself}. 
	
	For each $k \in 3, \dots, K,$ we can apply the following expansion to the odds ratio term. Without loss of generality we drop fully observed random variables $O$ for brevity,
	{\small
		\begin{align}
			\text{OR}(R_k, R_{\prec k} \mid R_{\succ k}=1, X^{(1)}) = \text{OR}(R_k, R_{k-1} \mid R_{-(k, k-1)}=1, X^{(1)}) \times \text{OR}(R_k, R_{\prec k-2} \mid R_{\succ k}=1, R_{k-1}, X^{(1)}).
			\label{eq:trick}
		\end{align}
	}
	This expansion can be applied inductively to the second term in the above product until $\text{OR}(R_k, R_{\prec k} \mid R_{\succ k}=1, X^{(1)})$ is expressed as a function of pairwise odds ratios and higher-order interaction terms. Applying the inductive expansion to each odds ratio term in $\prod_{k = 2}^{K} \text{OR}(R_k, R_{\prec k} \mid R_{\succ k} = 1, X^{(1)})$ we can re-express the identifying functional as,
	{\small
		\begin{align}
			p(R \mid X^{(1)}) = &\frac{1}{Z}\times \prod_{k = 1}^{K} \ p(R_k \mid R_{-k} = 1,X^{(1)})  \nonumber \\
			&\times \hspace{0.cm} \prod_{R_k, R_l \in R} \text{OR}(R_k, R_l \mid R_{-(k,l)}=1, X^{(1)}) \times \prod_{R_k, R_l, R_m \in R} f(R_k, R_l, R_m \mid R_{-(k, l, m)}=1, X^{(1)})  \nonumber\\
			&\times \hspace{-0.5cm}\prod_{R_k, R_l, R_m, R_n \in R} f(R_k, R_l, R_m, R_n \mid R_{-(k, l, m, n)}=1, X^{(1)}) \times \dots \times f(R_1, \dots, R_K \mid X^{(1)}),
			\label{eq:odds_proof}
		\end{align}
	}
	where $Z$ is the normalizing constant as before, and each $f(\cdot \mid \cdot, X^{(1)})$ are 3-way, 4-way, up to $K$-way interaction terms. 
	These interaction terms are defined as follows. 
	{\scriptsize
		\[f(R_i,R_j,R_k|R_{-(i,j,k)}=1,X^{(1)}) = \frac{\Odds(R_i,R_j|R_k, R_{-(i,j,k)}=1,X^{(1)})}{\Odds(R_i,R_j|R_k=1, R_{-(i,j,k)}=1,X^{(1)})},
		\] } and 
	{\scriptsize
		\[f(R_i,R_j,R_k,R_l|R_{-(i,j,k,l)}=1,X^{(1)}) = \frac{\Odds(R_i,R_j|R_k, R_l, R_{-(i,j,k,l)}=1,X^{(1)})}{\Odds(R_i,R_j|R_k=1, R_l, R_{-(i,j,k,l)}=1,X^{(1)})} \times \frac{\Odds(R_i,R_j|R_k=1, R_l=1, R_{-(i,j,k,l)}=1,X^{(1)})}{\Odds(R_i,R_j|R_k, R_l=1, R_{-(i,j,k,l)}=1,X^{(1)})},
		\] } and so on, up to 
	{\scriptsize 
		\begin{align*}
			f(R_1,...,R_K \mid X^{(1)}) &=  \Odds(R_i,R_j|R_{-(i,j)},X^{(1)}) \\ 
			& \times \frac{ \displaystyle \prod_{R_k, R_l \in R}\Odds(R_i,R_j|R_{(k,l)}=1,R_{-(i,j,k,l)},X^{(1)}) \prod_{R_k, R_l, R_m, R_n \in R}\Odds(R_i,R_j|R_{(k,l,m,n)}=1,R_{-(i,j,k,l,m,n)},X^{(1)}) \times \cdots }{\displaystyle \prod_{R_k \in R} \Odds(R_i,R_j|R_k=1,R_{-(i,j,k)},X^{(1)}) \prod_{R_k, R_l, R_m \in R}\Odds(R_i,R_j|R_{(k,l,m)}=1,R_{-(i,j,k,l,m)},X^{(1)}) \times \cdots }.
		\end{align*}
	}
	
	Readers familiar with the clique potential factorization of Markov random fields may treat these interaction terms analogously \citep{malinsky19noself}. We now show that each term in the above factorization is identified.
	
	\vspace{0.5cm}
	\noindent \underline{\textbf{Step 1.}} \\
	We start off by looking at the conditional pieces $p(R_k \mid R_{-k} = 1, X^{(1)}, O)$. Given Lemma.~\ref{lem:chain_submodel_dag}, we know that $R_k \ci X^{(1)}_k \mid R_{-k}, X^{(1)}_{-k}, O$. Therefore, $p(R_k \mid R_{-k} = 1, X^{(1)}, O) = p(R_k \mid R_{-k} = 1, X^{(1)}_{-k}, O),  \forall k,$ is identified for all $R_k \in R.$
	
	\vspace{0.5cm}
	\noindent \underline{\textbf{Step 2.}} \\
	We now show that for any $R_k, R_l \in R$, the pairwise odds ratio $\text{OR}(R_k, R_l \mid R_{\{ - (k, l) \}} = 1, X^{(1)}) $ given in Eq.~(\ref{eq:odds_proof}) is identified. We know that 
	\[
	\text{OR}(R_k, R_l \mid R_{ - (k, l) } = 1, X^{(1)})  = \text{OR}(R_k, R_l \mid R_{ - (k, l) } = 1, X_{- (k, l) }, X^{(1)}_k, X^{(1)}_l). 
	\]
	Consequently, if we can show that the odds ratio is neither a function of $X^{(1)}_k$ nor $X^{(1)}_l$, then we can safely claim that the odds ratio is only a function of observed data and hence is identified. We get to this conclusion by exploiting the symmetric notion in odds ratios. 
	\begin{align*}
		\text{OR}(R_k, R_l \mid R_{- (k, l)} = 1, X^{(1)}) 
		&= \frac{p(R_k \mid R_l, R_{- (k, l) } = 1, X^{(1)})}{p(R_k =1 \mid R_l, R_{- (k, l) } = 1, X^{(1)})} \times \frac{p(R_k = 1 \mid R_{-k} = 1, X^{(1)})}{p(R_k \mid R_{-k} = 1, X^{(1)})} 
		\\
		&= \frac{p(R_l \mid R_k, R_{- (k, l)} = 1, X^{(1)})}{p(R_l =1 \mid R_k, R_{- (k, l) } = 1, X^{(1)})} \times \frac{p(R_l = 1 \mid R_{-l} = 1, X^{(1)})}{p(R_l \mid R_{-l} = 1, X^{(1)})} 
	\end{align*}
	In the first equality, we can see that the odds ratio is not a function of $X^{(1)}_k$ since $R_k \ci X^{(1)}_k \mid R_{-k}, X^{(1)}_{-k}$. Similarly, from the second equality, we can see that the odds ratio is not a function of $X^{(1)}_l$ since $R_l \ci X^{(1)}_l \mid R_{-l}, X^{(1)}_{-l}$. Therefore, the pairwise odds ratios are all identified.
	
	Finally we show that each of the higher-order interaction terms are identified. For each of these terms we need to show that they are not a function of missing variables with indices corresponding to indicators to the left of the conditioning bar. That is, we need to show that the 3-way interaction terms $f(R_k, R_l, R_m \mid R_{-(k, l, m)}=1, X^{(1)})$ are not functions of $X_{(k, l, m)}^{(1)},$ the 4-way interaction terms $f(R_k, R_l, R_m, R_n \mid R_{-(k, l, m, n)}=1, X^{(1)})$ are not functions of $X_{(k, l, m, n)}^{(1)}$, and so on until finally the $K$-way interaction term $f(R_1, \dots, R_K \mid X^{(1)})$ is not a function of $X^{(1)}.$ 
	
	Because of the way the odds ratio is defined, each $f(\cdot \mid \cdot, X^{(1)})$ is symmetric in the $k$ arguments appearing to the left of the conditioning bar and can be rewritten in multiple equivalent ways. In particular, each $k$-way interaction term can be rewritten in $\binom{k}{2}$ ways for any choice of indices $i, j$ of the missingness indicators that appear to the left of the conditioning bar. Each such representation allows us to conclude that $f(\cdot \mid \cdot, X^{(1)})$ is not a function of $X_i^{(1)}, X_j^{(1)}.$ Combining all these together allows us to conclude that the $k$-way interaction term $f(\cdot \mid \cdot, X^{(1)})$ is not a function of the missing variables corresponding to the indicators appearing on the left of the conditioning bar. 
	
	As a concrete example, consider the 3-way interaction $f(R_1, R_2, R_3 \mid R_{-(1, 2, 3)} = 1, X^{(1)}).$ We can write it down in three different ways as follows. 
	{\scriptsize
		\begin{align*}
			&f(R_i, R_j, R_k  \mid R_{-(1, 2, 3)} = 1, X^{(1)}) 
			\\
			&= \frac{\text{OR}(R_1, R_2 \mid R_{-(1, 2, 3)} = 1, R_3, X^{(1)}) }{\text{OR}(R_1, R_2 \mid R_{-(1, 2, 3)} = 1, R_3 = 1, X^{(1)})}  
			= \frac{\text{OR}(R_1, R_3 \mid R_{-(1, 2, 3)} = 1, R_2, X^{(1)}) }{\text{OR}(R_1, R_3 \mid R_{-(1, 2, 3)} = 1, R_2 = 1, X^{(1)})} 
			= \frac{\text{OR}(R_2, R_3 \mid R_{-(1, 2, 3)} = 1, R_1, X^{(1)}) }{\text{OR}(R_2, R_3 \mid R_{-(1, 2, 3)} = 1, R_1 = 1, X^{(1)})}
		\end{align*}
	}%
	From the first equality, we note that $f$ is not a function of $X^{(1)}_1, X^{(1)}_2$. From the second equality, we note that $f$ is not a function of $X^{(1)}_1, X^{(1)}_3$. From the third equality, we note that $f$ is not a function of $X^{(1)}_2, X^{(1)}_3$. Therefore, $f$ is not a function of $X^{(1)}_1, X^{(1)}_2, X^{(1)}_3$ and is identified. 
\end{proof}

\begin{thma}{\ref{theorem:sound_complete_dag}}
	The graphical condition of no self-censoring and no colluders, put forward in Theorem~\ref{theorem:full_law_dag}, is sound and complete for the identification of full laws $p(R, O, X^{(1)})$ that are Markov relative to a missing data DAG $\G.$
\end{thma}
\begin{proof}
	Soundness is a direct consequence of Theorem~\ref{theorem:full_law_dag}. To prove completeness, it needs to be shown that in the presence of a self-censoring edge, or a colluder structure, the full law is no longer (non-parametrically) identified. A proof by counterexample of both these facts was provided in \citep{rozi19mid}. However, this can also be seen from the fact that self-censoring edges and colluders are special cases of the colluding paths that we prove results in non-identification of the full law in Lemma~\ref{lem:colluding_paths_nonid}.
\end{proof}

\begin{thma}{\ref{theorem:full_law_admg}}
	A full law $p(R, X^{(1)}, O)$ that is Markov relative to a missing data ADMG $\G$ is identified if  $\G$ does not contain any colluding paths and the stated positivity assumption in Section~\ref{sec:full_law_DAG} holds. Moreover, the resulting identifying functional for the missingness mechanism $p(R \mid X^{(1)}, O)$ is given by the odds ratio parametrization provided in Eq.~\ref{eq:odds_ratio_chen} of the main draft. 
\end{thma}
\begin{proof}
	The proof strategy is nearly identical to the one utilized in Theorem~\ref{theorem:full_law_dag}, except the conditional independences $R_k \ci X_k^{(1)} \mid R_{-k}, X_{-k}^{(1)}, O$ come from Lemma~\ref{lem:chain_submodel_admg} instead of Lemma~\ref{lem:chain_submodel_dag}.
\end{proof}

\begin{lema}{\ref{lem:colluding_paths_nonid}}
	A full law $p(R, X^{(1)}, O)$ that is Markov relative to a missing data ADMG $\G$ containing a colluding path between any pair $X_i^{(1)} \in X^{(1)}$ and $R_i \in R,$ is not identified.
\end{lema}

\begin{figure}[t]
	\begin{center}
		\scalebox{0.7}{
			\begin{tikzpicture}[>=stealth, node distance=2cm]
			\tikzstyle{format} = [thick, circle, minimum size=1.0mm,
			inner sep=0pt]
			\begin{scope}[xshift=0cm]
			\path[->, very thick]
			node[format] (x11) {$X^{(1)}_1$}
			node[format, below of=x11] (r1) {$R_1$}	
			node[format, below of=r1] (x1) {$X_1$}
			(x11) edge[red, <->] (r1)
			(r1) edge[gray] (x1)
			(x11) edge[gray, bend right] (x1)
			node[below of=x1, yshift=0.75cm, xshift=0cm] {(a)} 
			;
			\end{scope} 
			\begin{scope}[xshift=2cm]
			\path[->, very thick]
			node[format] (x11) {$X^{(1)}_1$}
			node[format, below of=x11, yshift=1cm] (h) {$U$}
			node[format, below of=x11] (r1) {$R_1$}	
			node[format, below of=r1] (x1) {$X_1$}
			(h) edge[blue] (r1)
			(h) edge[blue] (x11)
			(r1) edge[gray] (x1)
			(x11) edge[gray, bend right] (x1)
			node[below of=x1, yshift=0.75cm, xshift=0cm] {(b)} 
			;
			\end{scope} 
			\begin{scope}[xshift=4cm]
			\path[->, very thick]
			node[format] (x11) {}
			node[format, below of=x11] (r1) {$R_1$}	
			node[format, below of=r1] (x1) {$X_1$}
			(r1) edge[gray] (x1)
			(r1) edge[red, <->, bend right] (x1)
			node[below of=x1, yshift=0.75cm, xshift=0cm] {(c)} 
			;
			\end{scope} 
			\begin{scope}[xshift=7.5cm]
			\path[->, very thick]
			node[format] (x11) {$X^{(1)}_1$}
			node[format, right of=x11] (x21) {$X^{(1)}_2$}
			node[format, below of=x11] (r1) {$R_1$}		
			node[format, below of=x21] (r2) {$R_2$}
			node[format, below of=r1] (x1) {$X_1$}
			node[format, below of=r2] (x2) {$X_2$}
			(x11) edge[blue] (x21)
			(r1) edge[red, <->] (x21)
			(r1) edge[gray] (x1)
			(r2) edge[gray] (x2)
			(x11) edge[gray, bend right] (x1)
			(x21) edge[gray, bend left] (x2)
			node[below of=x1, yshift=0.75cm, xshift=1cm] {(d)} 
			;
			\end{scope} 
			\begin{scope}[xshift=11.5cm]
			\path[->, very thick]
			node[format] (x11) {$X^{(1)}_1$}
			node[format, right of=x11] (x21) {$X^{(1)}_2$}
			node[format, below of=x11] (r1) {$R_1$}		
			node[format, below of=x21] (r2) {$R_2$}
			node[format, below of=r1] (x1) {$X_1$}
			node[format, below of=r2] (x2) {$X_2$}
			(x11) edge[red, <->] (x21)
			(r1) edge[red, <->] (x21)
			(r1) edge[gray] (x1)
			(r2) edge[gray] (x2)
			(x11) edge[gray, bend right] (x1)
			(x21) edge[gray, bend left] (x2)
			node[below of=x1, yshift=0.75cm, xshift=1cm] {(e)} 
			;
			\end{scope} 
			\begin{scope}[xshift=15cm]
			\path[->, very thick]
			node[format] (x11) {}
			node[format, right of=x11] (x21) {}
			node[format, below of=x11] (r1) {$R_1$}		
			node[format, below of=x21] (r2) {$R_2$}
			node[format, below of=r1] (x1) {$X_1$}
			node[format, below of=r2] (x2) {$X_2$}
			(r1) edge[red, <->] (x2)
			(x1) edge[red, <->] (x2)
			(r1) edge[gray] (x1)
			(r2) edge[gray] (x2)
			node[below of=x1, yshift=0.75cm, xshift=1cm] {(f)} 
			;
			\end{scope} 
			\end{tikzpicture}
		}
	\end{center}
	\caption{(a, d, e) Examples of colluding paths in missing data models of ADMGs. (b) A DAG with hidden variable $U$ that is Markov equivalent to (a).  (c) Projecting out $X^{(1)}_1$ from (a), (f) Projecting out $X^{(1)}_1$ and $X^{(1)}_2$ from (d) and (e). }
	\label{fig:colluding_ex_1}
\end{figure}
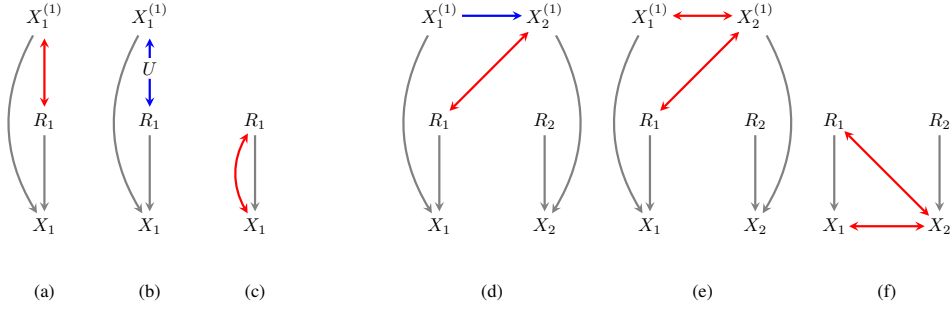

\begin{proof}
	
	Proving the non-identifiability of missing data models of an ADMG $\G$ that contains a colluding path can be shown by providing two models ${\cal M}_1$ and ${\cal M}_2$ that disagree on the full law but agree on the observed law. Coming up with a single example of such a pair of models is sufficient for arguing against non-parametric identification of the full law. Therefore, for simplicity, we restrict our attention to binary random variables. We first provide an example of such a pair of models on the simplest form of a colluding path, a bidirected edge $X_i^{(1)} \leftrightarrow R_i$ as shown in Fig.~\ref{fig:colluding_ex_1}(a). According to Table~\ref{table:counter_example}, in order for the observed laws to agree, the only requirement is that the quantity $ab + (1-a)c$ remain equal in both models; hence we can come up with infinitely many counterexamples of full laws that are not the same but map to the same observed law.
	
	Constructing explicit counterexamples are not necessary to prove non-identification as long as it can be shown that there exist at least two distinct functions that map two different full laws onto the exact same observed law. For instance, if the number of parameters in the full law is strictly larger than the number of parameters in the observed law, then there  would exist infinitely many such functions. Consequently, we rely on a parameter counting argument to prove the completeness of our results. Since we are considering missing data models of ADMGs, we use the Moebius parameterization of binary nested Markov models of an ADMG described in Appendix A. 
	
	The nested Markov model of a missing data ADMG $\G(V),$ where $V=\{O,X^{(1)}, R, X\},$ is a smooth super model of the missing data DAG model $\G(V\cup U),$ and has the same model dimension as the latent variable model \citep{evans2018margins}. We also utilize nested Markov models of an ADMG $\G(V\setminus X^{(1)}),$ corresponding to projection of the missing data ADMG $\G(V)$ onto variables that are fully observable. While such a model does not capture all equality constraints in the true observed law, it is still a smooth super model of it, thus providing an \emph{upper bound} on the model dimension of the observed law. This suffices for our purposes, as demonstrating that the upper bound on the number of parameters in the observed law is less than the number of parameters in the full law, is sufficient to prove that the full law is not identified. We first walk the reader through a few examples to demonstrate this proof strategy, and then provide the general argument.
	
	\begin{table}[t]
		\begin{center}
			\scalebox{0.88}{
				\begin{tabular}{| c | c |}
					\hline
					$U$ & $p(U)$ \\ \hline
					$0$  & $a$     \\ 
					$1$  & $1-a$  \\ \hline
				\end{tabular}
				
				\hspace{0.5cm}
				
				\begin{tabular}{| c : c | c |}
					\hline
					$R_1$ & $U$ & $p(R_1 | U)$ \\ \hline
					$0$  & $0$ & $b$  \\ 
					$1$  & $0$ & $1 - b$ \\ \hline
					$0$  & $1$ & $c$  \\ 
					$1$  & $1$ & $1 - c$ \\ 
					\hline
				\end{tabular}
				
				\hspace{0.5cm}
				
				\begin{tabular}{| c : c | c |}
					\hline
					$X^{(1)}_1$ & $U$ & $p(X^{(1)}_1 | U)$ \\ \hline
					$0$  & $0$ & $d$  \\ 
					$1$  & $0$ & $1 - d$ \\ \hline
					$0$  & $1$ & $e$  \\ 
					$1$  & $1$ & $1 - e$ \\ 
					\hline
				\end{tabular}
				
				\hspace{0.5cm} 
				
				\begin{tabular}{| c c c | c |}
					\hline
					$R_1$ & $X^{(1)}_1$ & $U$ & $p(R_1, X^{(1)}_1, U)$ \\ \hline
					$0$  & $0$ & $0$ & $a*b*d$  \\ 
					$0$  & $0$ & $1$ & $(1-a)*c*e$  \\ \hline 
					$0$  & $1$ & $0$ & $a*b*(1-d)$  \\ 
					$0$  & $1$ & $1$ & $(1-a)*c*(1-e)$  \\ \hline 
					$1$  & $0$ & $0$ & $a*(1-b)*d$  \\ 
					$1$  & $0$ & $1$ & $(1-a)*(1-c)*e$  \\ \hline 
					$1$  & $1$ & $0$ & $a*(1-b)*(1-d)$  \\ 
					$1$  & $1$ & $1$ & $(1-a)*(1-c)*(1-e)$  \\ 
					\hline
				\end{tabular}
			}
			\scalebox{0.88}{
				\begin{tabular}{ | c | c | c | c | c | }
					\hline
					$R_1$  & $X^{(1)}_1$  & p(Full Law)  & $X_1$ &  p(Observed Law)    \\ \hline
					\multirow{2}{*}{0} &  0   &  $a*b*d + (1-a)*c*e$   & \multirow{2}{*}{?} & \multirow{2}{*}{$a*b + (1-a)*c$}   \\ 
					& 1  &  $a*b*(1-d) + (1-a)*c*(1-e)$  &  &    \\ 
					
					\hline
					
					\multirow{2}{*}{1} &  0   &  $a*(1-b)*d + (1-a)*(1-c)*e$   & 0 & \multirow{2}{*}{$a*(1-b) + (1-a)*(1-c)$}   \\ 
					& 1  &  $a*(1-b)*(1-d) + (1-a)*(1-c)*(1-e)$  & 1 &    \\ \hline 
					
				\end{tabular}
			}
			\caption{Construction of counterexamples for non-identifiablity of the full law in Fig.~\ref{fig:colluding_ex_1}(a) using the DAG with hidden variable $U$ in Fig.~\ref{fig:colluding_ex_1}(b) that is Markov equivalent to (a). }
			\label{table:counter_example}
		\end{center}
	\end{table}
	
	\begin{table}[t]
		\begin{center}
			\scalebox{1}
			{
				\begin{tabular}{ | c | c | c | c | }
					\hline
					\multicolumn{4}{|c|}{\multirow{2}{*}{Moebius Parameterization of the Full Law in Fig.~\ref{fig:colluding_ex_1}(d)}} \\ 
					\multicolumn{4}{|c|}{}\\ \hline
					Districts  & Intrinsic Head/Tail  & Moebius Parameters  & Counts    \\ \hline
					$\{X^{(1)}_1\}$ & $\{X^{(1)}_1\}, \{ \}$ & $q(X^{(1)}_1 = 0)$ & $1$ \\ \hline 
					$\{R_2\}$ & $\{R_2\}, \{ \}$ & $q(R_2 = 0)$ & $1$ \\ \hline 
					\multirow{2}{*}{$\{R_1, X^{(1)}_2\}$} & $\{R_1\}, \{\}$ & $q(R_1 = 0)$ & $1$ \\ 
					& $\{X^{(1)}_2\}, \{X^{(1)}_1\}$ & $q(X^{(1)}_2 = 0 \mid X^{(1)}_1)$ & $2$ \\ 
					& $\{R_1, X^{(1)}_2\}, \{X^{(1)}_1\}$ & $q(R_1 = 0, X^{(1)}_2 = 0 \mid X^{(1)}_1)$ & $2$ \\  \hline
					\multicolumn{3}{|r|}{Total} & $7$ \\
					\multicolumn{3}{|r|}{} &  \\ 
					\hline 
					\multicolumn{4}{|c|}{\multirow{2}{*}{Moebius Parameterization of the Full Law in Fig.~\ref{fig:colluding_ex_1}(e)}} \\ 
					\multicolumn{4}{|c|}{}\\ \hline
					Districts  & Intrinsic Head/Tail  & Moebius Parameters  & Counts    \\ \hline
					$\{R_2\}$ & $\{R_2\}, \{ \}$  & $q(R_2 = 0)$ & $1$ \\ \hline
					\multirow{6}{*}{$\{R_1, X^{(1)}_1, X^{(1)}_2\}$} & $\{R_1\}, \{\}$ & $q(R_1 = 0)$ & $1$ \\ 
					& $\{X^{(1)}_1\}, \{ \}$ & $q(X^{(1)}_1 = 0)$ & $1$ \\ 
					& $\{X^{(1)}_2\}, \{ \}$ & $q(X^{(1)}_2 = 0)$ & $1$ \\ 
					& $\{R_1, X^{(1)}_2\}, \{ \}$ & $q(R_1 = 0, X^{(1)}_2 = 0)$ & $1$ \\ 
					& $\{X^{(1)}_1, X^{(1)}_2\}, \{ \}$ & $q(X^{(1)}_1 = 0, X^{(1)}_2 = 0)$ & $1$ \\ 
					& $\{R_1, X^{(1)}_1, X^{(1)}_2\}, \{ \}$ & $q(R_1 = 0, X^{(1)}_1 = 0, X^{(1)}_2 = 0)$ & $1$ \\ \hline
					\multicolumn{3}{|r|}{Total} & $7$ \\
					\multicolumn{3}{|r|}{} &  \\ 
					\hline 
					\multicolumn{4}{|c|}{\multirow{2}{*}{Moebius Parameterization of the Observed Law in Fig.~\ref{fig:colluding_ex_1}(f)}} \\ 
					\multicolumn{4}{|c|}{}\\ \hline
					Districts  & Intrinsic Head/Tail  & Moebius Parameters  & Counts    \\ \hline
					$R_2$ & $\{R_2\}, \{\}$ & $q(R_2 = 0)$ & $1$ \\ \hline
					\multirow{5}{*}{$\{R_1, X_1, X_2\}$} & $\{R_1\}, \{\}$ & $q(R_1 = 0)$ & $1$ \\ 
					& $\{X_1\}, \{R_1 \}$ & $q(X_1 = 0 \mid R_1)$ & $1$ \\ 
					& $\{X_2\}, \{R_2\}$ & $q(X_2 = 0 \mid R_2)$ & $1$ \\ 
					& $\{R_1, X_2\}, \{R_2\}$ & $q(R_1 = 0, X_2 = 0 \mid R_2)$ & $1$ \\ 
					& $\{X_1, X_2\}, \{R_1, R_2\}$ & $q(X_1 = 0, X_2 = 0 \mid R_1, R_2)$ & $1$ \\  \hline 
					\multicolumn{3}{|r|}{Total} & $6$ \\ 
					\multicolumn{3}{|r|}{} &  \\ 
					\hline
				\end{tabular}
			}
			\caption{Moebius Parameterization of the Full and Observed Laws of missing data ADMGs}
			\label{tab:moebius_counts}
		\end{center}
	\end{table}

	\subsubsection*{Self-censoring through unmeasured confounding:}
	
	We start by reanalyzing the colluding path given in Fig.~\ref{fig:colluding_ex_1}(a) and the corresponding projection given in Fig.~\ref{fig:colluding_ex_1}(c). The Moebius parameters associated with the full law are $q(X^{(1)}_1 = 0), q(R_1 = 0), q(X^{(1)}_1 = 0, R_1 = 1),$ for a total of 3 parameters. The Moebius parameters associated with the observed law in Fig~\ref{fig:colluding_ex_1}(c) are $q(R_1=0), q(X^{(1)}_1= 0 \mid R_1 = 0),$ for a total of only 2 parameters. Since $2 < 3$, we can construct infinitely many mappings, as it was shown in Table~\ref{table:counter_example}. 
	
	\subsubsection*{Simple colluding paths:}
	
	Consider the colluding paths given in Fig.~\ref{fig:colluding_ex_1}(d, e) and the corresponding projection (which are identical in both cases) given in Fig.~\ref{fig:colluding_ex_1}(f). The Moebius parameters associated with the full laws and observed law are shown in Table~\ref{tab:moebius_counts}. Once again, since the number of parameters in the observed law is less than the number in the full law $(6 < 7)$, we can construct infinitely many mappings. 
	
	\clearpage
	\subsubsection*{A general argument:}
	
	In order to generalize our argument, we first provide a more precise representation (that does not use dashed edges) in Figs.~\ref{fig:proof}(a-d), of all possible colluding paths between $X_i^{(1)}$ and $R_i.$ Without loss of generality, assume that there are $K$ variables in $X^{(1)}$ and there are $S$ variables that lie on the collider path between $X^{(1)}_i$ and $R_i$, $S \in \{0, 1, \dots, 2*(K-1)\}$. We denote the $s$th variable on the collider path by $V_s$; $V_s \in \{X^{(1)} \setminus X^{(1)}_i, R \setminus R_i\}.$ Note that $V_S$ in Figs.~\ref{fig:proof}(c, d) can only belong to $\{R \setminus R_i\}$ by convention. Fig.~\ref{fig:proof}(e) illustrates the corresponding projections of figures (a) and (b), and Fig.~\ref{fig:proof}(f) illustrates the corresponding projections of figures (c) and (d). In the projections shown in Figs.~\ref{fig:proof}(e, f), $V^* \in \{X \setminus X^{(1)}_i, R \setminus R_i\}.$
	
	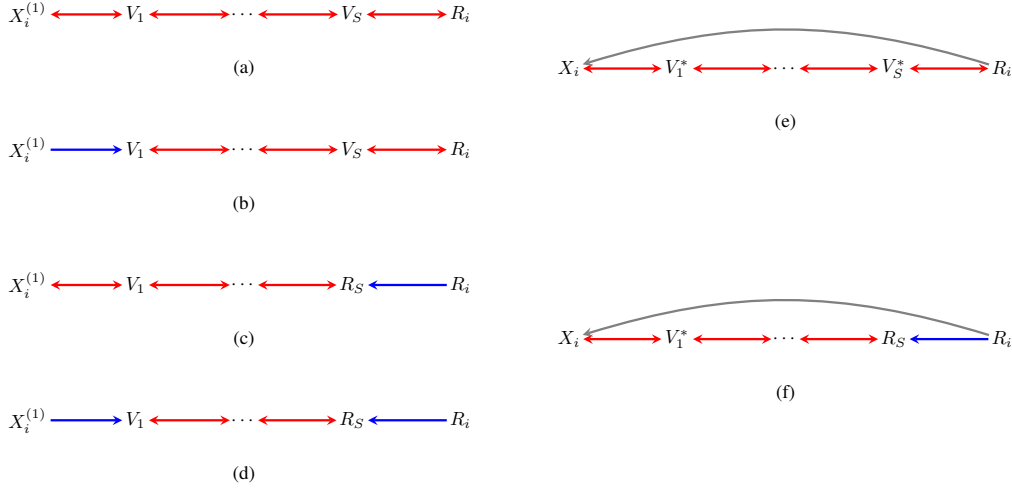
\begin{figure}[t]
		\begin{center}
			\scalebox{0.72}{
				\begin{tikzpicture}[>=stealth, node distance=2cm]
				\tikzstyle{format} = [thick, circle, minimum size=1.0mm, inner sep=0pt]
				\tikzstyle{square} = [draw, thick, minimum size=1mm, inner sep=3pt]	
				\begin{scope}[xshift=0cm]
				\path[->, very thick]
				node[format] (xi) {$X^{(1)}_i$}
				node[format, right of=xi] (rj) {$V_1$}
				node[format, right of=rj] (dot) {$\cdots$}
				node[format, right of=dot] (rk) {$V_S$}
				node[format, right of=rk] (ri) {$R_i$}
				
				(xi) edge[red, <->] (rj)
				(rj) edge[red, <->] (dot)
				(dot) edge[red, <->] (rk)
				(ri) edge[red, <->] (rk)
				node[below of=dot, yshift=1cm, xshift=0cm] {(a)} 
				;
				\end{scope}
				\begin{scope}[xshift=0cm, yshift=-2.5cm]
				\path[->, very thick]
				node[format] (xi) {$X^{(1)}_i$}
				node[format, right of=xi] (rj) {$V_1$}
				node[format, right of=rj] (dot) {$\cdots$}
				node[format, right of=dot] (rk) {$V_S$}
				node[format, right of=rk] (ri) {$R_i$}
				
				(xi) edge[blue] (rj)
				(rj) edge[red, <->] (dot)
				(dot) edge[red, <->] (rk)
				(ri) edge[red, <->] (rk)
				node[below of=dot, yshift=1cm, xshift=0cm] {(b)} 
				;
				\end{scope}
				\begin{scope}[xshift=0cm, yshift=-5cm]
				\path[->, very thick]
				node[format] (xi) {$X^{(1)}_i$}
				node[format, right of=xi] (xj) {$V_1$}
				node[format, right of=xj] (dot) {$\cdots$}
				node[format, right of=dot] (rk) {$R_S$}
				node[format, right of=rk] (ri) {$R_i$}
				
				(xi) edge[red, <->] (xj)
				(xj) edge[red, <->] (dot)
				(dot) edge[red, <->] (rk)
				(ri) edge[blue] (rk)
				node[below of=dot, yshift=1cm, xshift=0cm] {(c)} 
				;
				\end{scope}
				\begin{scope}[xshift=0cm, yshift=-7.5cm]
				\path[->, very thick]
				node[format] (xi) {$X^{(1)}_i$}
				node[format, right of=xi] (xj) {$V_1$}
				node[format, right of=xj] (dot) {$\cdots$}
				node[format, right of=dot] (rk) {$R_S$}
				node[format, right of=rk] (ri) {$R_i$}
				
				(xi) edge[blue] (xj)
				(xj) edge[red, <->] (dot)
				(dot) edge[red, <->] (rk)
				(ri) edge[blue] (rk)
				node[below of=dot, yshift=1cm, xshift=0cm] {(d)} 
				;
				\end{scope}
				\begin{scope}[xshift=10cm, yshift=-1.0cm]
				\path[->, very thick]
				node[format] (xi) {$X_i$}
				node[format, right of=xi] (rj) {$V^*_1$}
				node[format, right of=rj] (dot) {$\cdots$}
				node[format, right of=dot] (rk) {$V^*_S$}
				node[format, right of=rk] (ri) {$R_i$}
				
				(xi) edge[red, <->] (rj)
				(rj) edge[red, <->] (dot)
				(dot) edge[red, <->] (rk)
				(ri) edge[red, <->] (rk)
				(ri) edge[gray, bend right=17] (xi)
				node[below of=dot, yshift=1cm, xshift=0cm] {(e)} 
				;
				\end{scope}
				\begin{scope}[xshift=10cm, yshift=-6cm]
				\path[->, very thick]
				node[format] (xi) {$X_i$}
				node[format, right of=xi] (xj) {$V^*_1$}
				node[format, right of=xj] (dot) {$\cdots$}
				node[format, right of=dot] (rk) {$R_S$}
				node[format, right of=rk] (ri) {$R_i$}
				
				(xi) edge[red, <->] (xj)
				(xj) edge[red, <->] (dot)
				(dot) edge[red, <->] (rk)
				(ri) edge[blue] (rk)
				(ri) edge[gray, bend right=17] (xi)
				node[below of=dot, yshift=1cm, xshift=0cm] {(f)} 
				;
				\end{scope}	
				\end{tikzpicture}
			}
		\end{center}
		\caption{(a) Colluding paths (b) Projecting out $X^{(1)}$}
		\label{fig:proof}
	\end{figure}
	
	We now go over each of these colluding paths and their corresponding latent projections, as if they appear in a larger graph that is otherwise completely disconnected. We count the number of Moebius parameters as a function of $S$, and show that the full law always has one more parameter than the observed law. One can then imagine placing these colluding paths in a larger graph with arbitrary connectivity, and arguing that the full law is still not identified as a consequence of the parameter discrepancy arising from the colluding path alone. That is, if we show a fully disconnected graph containing a single colluding path is not identified, then it is also the case that any edge super graph (super model) is also not identified.
	
	In the following proof we heavily rely on the following fact. Given a bidirected chain of length $V_1 \leftrightarrow, \cdots, \leftrightarrow V_K,$ of length $K,$ the number of Moebius parameters required to parameterize this chain is given by the sum of natural numbers $1$ to $K,$ i.e., $\frac{K(K+1)}{2}.$ This can be seen from the fact that the corresponding Moebius parameters are given by the series,
	\begin{itemize}
		\item $q(V_1=0), q(V_1=0, V_2=0), \dots, q(V_1=0, \dots, V_K=0)$ corresponding to $K$ parameters.
		\item $q(V_2=0), q(V_2, V_3=0), \dots, q(V_2=0, \dots, V_K=0)$ corresponding to $K-1$ parameters.
		\item $\dots$
		\item $q(V_K=0)$ corresponding to $1$ parameter.
	\end{itemize}
	
	In counting the number of parameters for a disconnected graph (with the exception of the colluding path), we can also exclude the singleton (disconnected) nodes from the counting argument since they account for the same number of parameters in both the full law and observed law. In the full law they are either $q(R_s = 0)$ or $q(X^{(1)}_s = 0)$ and the corresponding parameters in the observed law are $q(R_s = 0)$ or $q(X_s = 0 \mid R_s = 1)$. The Moebius parameter counts for each of the colluding paths in Figs.~\ref{fig:proof}(a-d) and their corresponding latent projections in Figs.~\ref{fig:proof}(e,f) are as follows.
	
	\underline{\textbf{Figures a, b, and e}}
	\begin{enumerate}
		\item Number of Moebius parameters in Fig.~\ref{fig:proof}(a) is $\frac{(S+2)(S+3)}{2}$ 
		\begin{itemize}
			\item 
			A bidirected chain $X_i^{(1)} \leftrightarrow, \cdots, \leftrightarrow R_i$ of length $S+2$, i.e., $(S+2)*(S+3)/2$ parameters.
		\end{itemize}
		
		\item Number of Moebius parameters in Fig.~\ref{fig:proof}(b) is $\frac{(S+2)(S+3)}{2}$ 
		\begin{itemize}
			\item $q(X^{(1)}_i = 0)$, i.e. $1$ parameter,
			\item A bidirected chain $V_2 \leftrightarrow \cdots \leftrightarrow R_i$ of length $S$, i.e. $S*(S+1)/2$ parameters,
			\item Intrinsic sets involving $V_1$, i.e., $q(V_1=0 \mid X_i^{(1)}), q(V_1=0, V_2=0 \mid X_i^{(1)}), q(V_1=0, \dots, R_i=0 \mid X_i^{(1)})$ corresponding to $2*(S+1)$ parameters.
		\end{itemize}
		
		\item Number of Moebius parameters in Fig.~\ref{fig:proof}(e) is $\frac{(S+2)(S+3)}{2} - 1$ 
		\begin{itemize}
			\item Note that even though each proxy $X_s$ that may appear in the bidirected chain has a directed edge from $R_s$ pointing into it, the corresponding intrinsic head tail pair that involves both variables, will always have $R_i=1.$ Hence, we may ignore these deterministic edges and count the parameters as if it were a bidirected chain $V_1^* \leftrightarrow \cdots \leftrightarrow R_i$ of length $S+1$, corresponding to $(S+1)*(S+2)/2$ parameters,
			\item When enumerating intrinsic sets involving $X_i,$ we note that \textcolor{blue}{$\{X_i, V_1^*, \dots V_S^*\}$ is not intrinsic} as $R_i$ is not fixable (due to the bidirected path between $R_i$ and $X_i$ and the edge $R_i \rightarrow X_i$). Thus, as there is one less intrinsic set involving $X_i,$ the number of parameters required to parameterize all intrinsic sets involving $X_i$ is one fewer, i.e., $S+1$ (instead of $S+2$) parameters.
		\end{itemize}
	\end{enumerate}
	
	\underline{\textbf{Figures c, d, and f}}
	\begin{enumerate}
		\item Number of Moebius parameters in Fig.~\ref{fig:proof}(c) is $\frac{(S+2)(S+3)}{2}$ 
		\begin{itemize}
			\item $q(R_i = 0)$, i.e. $1$ parameter,
			\item A bidirected chain $X_i^{(1)}\leftrightarrow \cdots \leftrightarrow V_{S-1}$ of length $S$, i.e. $S*(S+1)/2$ parameters,
			\item Intrinsic sets involving $R_S$, i.e., $q(R_S=0 \mid R_i),q(R_S=0, V_{S-1}=0 \mid R_i), \dots, q(R_S=0, V_{S-1}=0 \dots, X_i^{(1)} \mid R_i),$ corresponding to $2*(S+1)$ parameters.
		\end{itemize}
		
		\item Number of Moebius parameters in Fig.~\ref{fig:proof}(d) is $\frac{(S+2)(S+3)}{2}$ 
		\begin{itemize}
			\item $q(X^{(1)}_i = 0), q(R_i = 0)$, i.e. $2$ parameters,
			
			\item A bidirected chain $V_2 \leftrightarrow \cdots \leftrightarrow V_{S-2}$ of length $S-2$, i.e. $(S-2)*(S-1)/2$ parameters,
			
			\item Intrinsic sets involving $V_1$ and not $R_S,$  i.e., $q(V_1=0 \mid X_i^{(1)}), q(V_1=0, V_2=0 \mid X_i^{(1)}), \dots, q(V_1=0, V_2=0, \dots, V_{S-1} \mid X_i^{(1)}),$ corresponding to $2*(S-1)$ parameters,
			
			\item Intrinsic sets involving $R_S$ and not $V_1$, i.e., $q(R_S=0 \mid R_i), q(R_S=0, V_{S-1}=0 \mid R_i), \dots, q(R_S=0, V_{S-1}=0, \dots, V_2 \mid R_i)$ corresponding to $2*(S-1)$ parameters.
			
			\item The intrinsic set involving both $V_1$ and $R_S,$ i.e., $q(V_1=0, V_2=0, \dots, R_S=0 \mid X_i^{(1)}, R_i),$ corresponding to $4$ parameters.
		\end{itemize}
		
		\item Number of Moebius parameters in Fig.~\ref{fig:proof}(f) is $\frac{(S+2)(S+3)}{2} - 1$ 
		\begin{itemize}
			\item $q(R_i = 0)$, i.e. $1$ parameter,
			\item By the same argument as before, deterministic tails can be ignored. Hence, we have a bidirected chain $X_i \leftrightarrow \cdots \leftrightarrow V_{S-1}$ of length $S$, i.e. $S*(S+1)/2$ parameters,
			\item Intrinsic sets involving $R_S$, i.e., $q(R_S=0 \mid R_i), q(R_S=0, V_{S-1} \mid R_i), \dots, q(R_S, V_{S-1}, \dots, V_1 \mid R_i),$ corresponding to $2*S$ parameters, and the special intrinsic set which results in the observed law having one less parameter $\textcolor{blue}{q(R_S, V_{S-1}, \dots, V_1, X_i \mid R_i = 1)}$ corresponding to just $1$ parameter instead of 2 due to the presence of the proxy $X_i$ in the head and the corresponding $R_i$ in the tail.
		\end{itemize}
		
	\end{enumerate}
\end{proof}

\begin{thma}{\ref{theorem:sound_complete_admg}}
	The graphical condition of the absence of colluding paths, put forward in Theorem~\ref{theorem:full_law_admg}, is sound and complete for the identification of full laws $p(R, O, X^{(1)})$ that are Markov relative to a missing data ADMG $\G.$
\end{thma}
\begin{proof}
	Soundness is a direct consequence of Theorem~\ref{theorem:full_law_admg} and completeness is a direct consequence of Lemma.~\ref{lem:colluding_paths_nonid}. 
\end{proof}

\clearpage
\bibliography{references}
\bibliographystyle{icml2020}

\end{document}